\documentclass[12pt]{iopart}
\usepackage{iopams}  
\usepackage{mathtools}
\usepackage{graphicx}
\usepackage{dcolumn}
\usepackage{bm}
\usepackage[mathlines]{lineno}
\usepackage{xcolor}
\usepackage{graphicx, caption}
\usepackage{lipsum}
\usepackage{amsmath}
\usepackage[normalem]{ulem}
\usepackage{tikz}
\usepackage{amsthm}
\newtheorem{theorem}{Theorem}
\newtheorem{corollary}{Corollary}[theorem]
\newtheorem{definition}{Definition}
\begin{document}
\pagestyle{plain}
\title{Functional Renormalization analysis of Bose-Einstein Condensation through complex interaction in Harmonic Oscillator; Can Bendixson criteria be extended to complex time?}

\author{Vinayak M Kulkarni \textsuperscript{*} }

\address{Theoretical Sciences Unit, Jawaharlal Nehru Centre for Advanced Scientific Research,
Bangalore 560064, India.}
\ead{mallikarjun@jncasr.ac.in,vmkphysimath@gmail.com}

\vspace{10pt}

\begin{abstract}
An arbitrary form of complex potential perturbation in an oscillator consists of many exciting questions in open quantum systems. These often provide valuable insights in a realistic scenario when a quantum system interacts with external environments. Action renormalization will capture the phase of the wave functions; hence we construct wave function from Bethe ansatz and Frobenius methods. The unitary and non-unitary regimes are discussed to connect with functional calculations. We present a functional renormalization calculation for a  non-hermitian oscillator. A  dual space Left-Right formulation is worked out in functional bosonic variables to derive the flow equation for scale-dependent action. We show equivalence between vertex operator and permutation operators. The results can be compared with Wentzel–Kramers–Brillouin(WKB) calculation. We formally construct the Bosonic coherent states in the dual space;breaking symmetry will lead to anyonic coherent states. The limit cycle in renormalization trajectories for complex flow parameters, especially in extended, complex time limits indicating the need for revisiting the Bendixson theorem. 
\end{abstract}

\maketitle

\section{Introduction}
Quantum systems with complicated potential forms often require a sophisticated theory to access all the quantum mechanical model parameter regimes. Specifically, the dissipative systems need reformulated conventional quantum mechanics methods; for example, correlation functions must be defined in the appropriate fields. Although the functional renormalization(FRG) calculations are known to be intractable, these provide enough insights into the problem. The renormalization group theory developed by Wilson and  Kadanoff is non-perturbative, and it is constructed on effective field theories by integrating out short-distance fluctuations below a certain cut-off scale \cite{huggett1995renormalisation,wilson1971renormalization,wilson1974renormalization,wilson1983renormalization,bain2013effective,teller1989infinite,castellani2002reductionism}. 
Quasiperiodic dissipative systems studied by various people using the FRG calculations \cite{quasiperiodicity,ostlund1983universal}. Several interesting potential form is chosen in the dissipative oscillator models earlier \cite{SHO1,SHO2,SHO3,SHO4,SHO5,SHO6,SHO7}. The recent FRG calculations on non-Hermitian models from local potential approximation and new techniques to solve the Wetterich equation or resulting flow equation with various regulator choice has been studied \cite{bender2018asymptotic,bender2021pt}. The critical phenomena and field-theoretical problems addressed by Wetterich \cite{wetterich2001effective} from the average action method by Legendre transform and using the Wilson and Kadanoff ideas, the FRG is developed. These ideas in the dissipative system or especially the wavefunction renormalization \cite{machacek1983two} in non-hermitian systems should be modified due to the presence of singularity and defectiveness in diagonalizability of the two-point or higher-order correlation functions, which will also involve unitary and non-unitary time evolution. Hence, the FRG has to be implemented, so the calculations are more tractable even for the complicated form of the potentials, which is the motivation of this work.  These problems in a time-dependent scenario and out of equilibrium with external drive become extremely difficult to solve. An average Hamiltonian approach similar to the Wetterich has been attempted recently, where it is also discussed briefly for fermionic and bosonic Grassman integrals for action\cite{vacca2012functional,vacca2011functional}. We also use various FRG methods \cite{kopietz2010introduction,tarjus2008nonperturbative} attempted for quantum thermodynamics and dissipative systems.  \\
As a starting point, we compare our results with the non-Hermitian oscillator problems\cite{real,real1, ext,ext1} which are well understood and extensively studied from WKB, a complex extension of conventional methods, and some innovative quantization using contour integrals. The Bethe ansatz in the complex interaction case studied and derived the full wave function from comparing with the Frobenius solution of the complex oscillator. Also, we discussed the connection with the Gross-Pitavaski equation and when do we get the limit-cycle in the wavefunction. These ideas can also be extended to many-body problems, particularly in the renormalization context, for example, the flow of the vacuum energy's ground-state with dissipation and a choice of the regulator. Suppose we focus on the transition points for $N=1$ and $N=2$, where N is the power of the complex potential. In that case, we will introduce and derive a characteristic temperature scale for the considered Bosonic system. Surprisingly at the transition point, we found the power-law form of the temperature scale with the average boson number after the Matsubara sum. This power-law mimics Bose-Einstein condensation in the non-Hermitian system studied recently\cite{obs,obs1}. We also attempted to get a solution from the RG perspective for various nonlinear potentials and generalization to fractional powers. Coherent states are constructed for the complex extension to Grassmann(g) numbers to see the flow of these coherent states under dissipation.
It is necessary to build mathematical methods to deal with the complexity of the problem. Here we attempt how these open systems (which generally can have unitary and non-unitary regimes) can be treated in the RG context through certain mathematical tools and we compare with exact results.There exist various results in mathematics literature on limit cycles in the complex plane without intersecting the real axis\cite{khanedani1997first,ilyashenko2002centennial} which has a nice connection in physics with conservation of energy and time-reversal symmetries. Although the Poincare-Bendixson theorem is about the closed or periodic orbits in dynamical systems, there are no exact analogs in complex time that are important in the quantum mechanics or the path integral context. We explore these closed trajectories of wave functions or the action Real-Complex energy plane RG context with various complex flow parameters in space and time.
\section{Some Preliminaries}
\subsection{Integrability}
The integrability of the generic quantum problem in one dimension strongly depend on the boundary conditions at some points and potential form which is detailed in the article\cite{combot2018integrability},Where a theorem has proved for a integrable quantum potential say $\gamma$ comes from a generic function $M(z,c)=-\frac{\partial}{\partial z}\ln(z^b e^{az^2/2}\, _1F_1\left(e,w,az^2\right))$ or $M(z,c)=-\frac{\partial}{\partial z}\ln(z^b e^{f(z)}\, _1F_1\left(e,w,f(z)\right))$ in general except z all are constants which are related to the theorem 1 and 2 of the article\cite{combot2018integrability} later we get these solutions for the wave functions from various methods and we do have the following proposition to get the limit cycle in wave function real-imaginary plane. Note that the Hypergeometric functions are related together\cite{fatah2016confluent,takayama2003generating,beukers2007gauss} and in special limits can be expressed as product of incomplete gamma and gamma functions and they are analytic(expanded as series). We do have different corrections to these forms in later section but we don't claim these are general to any complex form of the potential. For one of the case we get $M(z,c)=-\frac{1}{z}+az+\,\frac{_1F'_1}{_1F_1} 2z^{1-b}$ which can be shown for all of the cases these are indeed a bethe ansatz form $\frac{_1F'_1}{_1F_1}=z^{b-1}(M(z,c)-az +\frac{1}{z})$.Any polynomial potential form can be expressed in the following,
\begin{equation}
    \begin{split}
        z^n=\frac{1+z^n-1+z^n}{2}=\frac{1}{2}(e^{\ln(1+z^n)}-e^{-\ln(1-z^n)})\\
        =(1+z^n) {_2F_1\big(\frac{1}{2},1,\frac{3}{2},-(1+z^n)^2\big)}-(1-z^n) {_2F_1\big(\frac{1}{2},1,\frac{3}{2},-(1-z^n)^2\big)}
    \end{split}
\end{equation}
Above show the multi valued function for the general n-polynomial potential as we discussed earlier the ansatz from such forms may or may not exhibit the limit cycles  but if we get these functions in the RG beta functions then this violates the c-theorem of RG.
\subsection{c and g Theorem in Renormallization}
Here we discuss and prove the c and g theorems respectively\cite{c_th,c_th1,g_th,g_th1} as following,Some of the consequences  of these theorems in non-Hermitian systems  are discussed here in the language of renormallization.All symbols have standard meaning otherwise they have defined in respective sections.
\begin{theorem}
Let  $c(\ g_1,\ g_2...)$ be the function of couplings $g_i \in \mathbb{R}$ where $i={1,2,..}$ then the real valued c function is always monotonically decreasing.$ \frac{d c}{dt}=\sum^n_{i=1}\frac{\partial g_i}{\partial t}\frac{\partial c}{\partial g_i}\leq 0 $
\end{theorem}
\begin{proof}
To prove $\frac{d C}{dt}=\sum^n_{i=1}\frac{\partial g_i}{\partial t}\frac{\partial C}{\partial g_i}\leq 0$ where t is some flow parameter.We have following definitions for 2D conformal field theory definitions for complex coordinates $x^2=z\bar{z}$
\begin{definition}
\begin{equation}
    \begin{split}
        C(g)=2z^4\langle T(x)T(0) \rangle|_{x^2=1} \\
        H_i(g)=z^2x^2\langle T(x)\Phi_i(0) \rangle|_{x^2=1}\\
        G_{ij}(g)=x^4\langle \Phi_i(x)\Phi_j(0) \rangle|_{x^2=1}
    \end{split}
\end{equation}
We can write a equivalent definition with a scale parameter $t=\log(z\bar{z})$ and new field expansion $\Theta=\beta^i(g)\Phi_i$ since every coupling g flow with the energy.
\begin{equation}
    \begin{split}
        \frac{F(t)}{z^4}=\langle T(x)T(0) \rangle \\
        \frac{H(t)}{z^3\bar{z}}=\langle T(x)\Theta(0) \rangle\\
        \frac{G(t)}{z^2\bar{z}^2}=\langle \Theta(x)\Theta(0) \rangle
    \end{split}
\end{equation}
\end{definition}
where all three F,G,  and H are the amplitudes and preserve the rptational symmetry.
Consider Zomolodchikov's c function to prove this ,
$c(g)=C(g)+4\beta^k H_k-6\beta^i\beta^jG_{ij}$ by definition we can use $\beta^i\partial_i C(g)=-6\beta^i H_i +2\beta^k\partial_k(\beta^iH_i)+\beta^j\beta^k\partial_kG_{ij}+\beta^j(\partial_i \beta^k)G_{jk}+\beta^j(\partial_j\beta^k)G_{ik}=\beta^k\partial_k H_i+\partial_i \beta^k H_k -H_i +2\beta^k G_{ik}$ Now from c function defined earlier and the proof is valid for 2D CFT only,
\begin{equation}
    \begin{split}
        \beta^k \partial_k c(g)&= \beta^k \partial_k C(g) +\beta^k \partial_k (\beta^i H_i)-6\beta^k\partial_k(\beta^i\beta^jG_{ij})\\
       & =[-6\beta^iH_i+6\beta^k\partial_k(\beta^iH_i)]-6\beta^k\partial_k(\beta^i\beta^jG_{ij})
    \end{split}
\end{equation}
Last term in the above can be expanded by taking through the differential operator and permuting the dummi indices $[j\to i,i\to k,k\to j]$ clockwise and in reverse order $[j\to k, k\to i,i\to j]$  
\begin{equation}
    \begin{split}
        \beta^k \partial_k c(g)=-12\beta^i\beta^jG_{ij} <0, \hspace{2mm} for \hspace{2mm} G_{ij}>0
    \end{split}
\end{equation}
\end{proof}
\begin{corollary}
The either of the beta functions and norm $G_{ij}$ are simultaneously reverse sign then the c-theorem can be proved as above.When only either of the $\beta^i$ and $G_{ij}$ reverse sign then c-theorem will be violated.Also if the $\beta^i$ are multi valued then also the theorem will be violated.
\begin{equation}
    \begin{split}
        \beta^k \partial_k c(g)=-12(\beta^1\beta^1G_{11}+\beta^2\beta^2G_{22}+...)
    \end{split}
\end{equation}
For traceless metric $G_{ij}$ it is shown that limit cycles exist for some analytically continued parameters\cite{limi}.
\end{corollary}
\begin{corollary}
Let a nth order generic beta function consist of complex coupling then we have the following Abel's type ODE as a RG equation,
\begin{equation}
    \begin{split}
        \frac{d(ge^{e^{i\theta})}}{d\log t}&=-\beta(g)\\
        &=-(g^2e^{2i\theta}+ g^3e^{3i\theta}...)\\
        &=-g^2e^{2i\theta}\sum_{n}(g^{n-1}e^{i(n-1)\theta})\\
        &=-\frac{g^2e^{2i\theta}(g^ne^{in\theta}-1)}{ge^{i\theta}-1}
    \end{split}
\end{equation}
Now if we separate the real and imaginary couplings we get the following
\begin{equation}
    \begin{split}
        \frac{d g_{real}}{d g_{im}}=\frac{d(g\cos{\theta})}{d(g\sin{\theta})}=(-1)^n\tan\bigg(2\theta+\arctan\big(\frac{g^n\sin{n\theta}}{g^n\cos{n\theta}-1}\big)+\arctan\big(\frac{g\sin{\theta}}{g\cos{\theta}-1}\big)\bigg)
    \end{split}
\end{equation}
from the above we can see the number of limit cycles for odd-n is equals RG loop order but for even-n there are no limit cycles. The c theorem in conventional perturbative RG violates for odd orders in imaginary couplings, irrespective of the couplings are multivalued or not.
\end{corollary}
\begin{corollary}
For complex scale $\log{t}$ as a flow parameter then we have,
\begin{equation}
    \begin{split}
        \frac{dg}{d\ln(t)}-i\frac{d g}{d\ln(e^{i\theta})}=\beta(g) 
    \end{split}
\end{equation}
Given the invariance the the flow parameter attains the fixed point irrespective of the order of $\beta$ function.
\begin{equation}
    \begin{split}
        t=ce^{-\theta}
    \end{split}
\end{equation}
\end{corollary}
\textbf{\textit{The above corollary 1.3 basically show that given the renormalizability it attains a spiral invarient in the imaginary and real scales.}}
\begin{corollary}
 Let's consider a general case when both coupling and flow parameter is comlpex numbers then we have the following,
 \begin{equation}
     \begin{split}
         \begin{pmatrix}
              \frac{\partial}{\partial \ln(t)} & -i\frac{\partial}{\partial \ln(e^{i\theta})}\\
              -i\frac{\partial}{\partial \ln(e^{i\theta})}& \frac{\partial}{\partial \ln(t)}
         \end{pmatrix}\begin{pmatrix}
              g_{real}\\g_{im}
         \end{pmatrix}=\begin{pmatrix}
              Re(\beta)\\Im(\beta)
         \end{pmatrix}
     \end{split}
 \end{equation}
 Let's call the the operator $\mathbb{T}$ on the couplings and if we do a uniary operation as the following,
 \begin{equation}
     \begin{split}
         UU^\dag \begin{pmatrix}
              \frac{\partial}{\partial \ln(t)} & -i\frac{\partial}{\partial \ln(e^{i\theta})}\\
              -i\frac{\partial}{\partial \ln(e^{i\theta})}& \frac{\partial}{\partial \ln(t)}
         \end{pmatrix} UU^\dag\begin{pmatrix}
              g_{real}\\g_{im}
         \end{pmatrix}=\begin{pmatrix}
              Re(\beta)\\Im(\beta)
         \end{pmatrix}\\
         \begin{pmatrix}
              \frac{\partial}{\partial \ln(t)}-i\frac{\partial}{\partial \ln(e^{i\theta})} & 0\\
              0& \frac{\partial}{\partial \ln(t)}+i\frac{\partial}{\partial \ln(e^{i\theta})}
         \end{pmatrix} \begin{pmatrix}
              g_{real}+g_{im}\\r_{real}-g_{im}
         \end{pmatrix}=\begin{pmatrix}
              Re(\beta)+Im(\beta)\\Re(\beta)-Im(\beta)
         \end{pmatrix}
     \end{split}
 \end{equation}
 Arbitrary many choices are possible for $U$ but we choose simplest one as $U=\frac{1}{\sqrt{2}}\begin{pmatrix}
      1&1\\1&-1
 \end{pmatrix}$ this basically indicate that we get the $\frac{\partial g_{real}}{\partial g_{im}}\frac{t\partial \theta}{\partial t}=\frac{Re(\beta)}{Im(\beta)}$. Hence irrespective of the loop order if the problem is Renormalizable then we always get the limit cycles in $Re(\beta)$ and $Im(\beta)$ plane. from above two corrolaries we have the following,
 \begin{equation}
     \begin{split}
         \frac{Re(\beta)}{Im(\beta)}=\frac{\partial g_{real}}{\partial g_{im}}(c) \because  t=ce^{-\theta}
     \end{split}
 \end{equation}
This shows in polar parametrization (one way to show that is express the Cartesian coordinates to polar as $\frac{d y}{d x}=\frac{dr sin(\theta)+rcos(\theta)}{dr cos(\theta)-rsin(\theta)}$ which basically gives as $r=e^{-\theta}$ form for equal real and imaginary part of the beta function.) \textbf{\textit{always we have log spiral in the couplings irrespective of the model.}} This we test in the various RG methods in following sections.
\end{corollary}
\begin{theorem}
Let $\beta$ is a boundary function and the entropy under a RG transformation  $\frac{\partial S}{\partial \beta}=-\beta\langle(\tilde{H}-\langle \tilde{H}\rangle)^2\rangle<0$, and for $\beta\neq0$ and H is any self-adjoint Hamiltonian. $\langle H \rangle \in \mathbb{R}$
\end{theorem}
\begin{proof}
If we show $\langle (\tilde{H}-\langle \tilde{H} \rangle)^2 \rangle >0$ then g theorem will be automatically verified. $\langle (\tilde{H}-\langle \tilde{H} \rangle)^2 \rangle =\langle (\tilde{H}^2+\langle \tilde{H} \rangle^2-2\tilde{H}\langle \tilde{H} \rangle \rangle=\langle \tilde{H}^2\rangle+\langle \tilde{H}\rangle^2-2\langle\tilde{H}\rangle\langle\tilde{H}\rangle=\langle \tilde{H}^2\rangle-\langle \tilde{H}\rangle^2$ which is nothing but the varience $\sigma_H$ and it is always positive. 
\end{proof}
\begin{corollary}
Let's consider a H which consist of all conjugate pairs of complex eigenvalues as following,

\begin{equation}
\begin{split}
    &H=\begin{pmatrix}\boldsymbol{\lambda}e^{i\theta}&\bold{0}\\
    \bold{0}&\boldsymbol{\lambda}e^{-i\theta}
\end{pmatrix}\\
 &\implies \sigma_{H}= Tr(H^2)-Tr(H)^2=\lambda^2(2\cos(2\theta)-4\cos(\theta)^2)\\
 &=-2\lambda^2 <0
\end{split}
\end{equation}
Hence whenever the mapped H under RG picks the diagonal form as above violates the g theorem. For a generic case of the non-hermitian matrices the variance can be shown as $\sigma_H=-2\lambda^2 e^{i\sum_l \theta_l}$ this shows g theorem can only hold for $\sum_l\theta_l=\pi$, also the inequality $\frac{Tr(H)}{n}\ge\det(H)^{1/n}$ does not hold and determinant can cross zero as well.
\end{corollary}

Let's define a non-linear transformation $\mathbb{T}$: transformation on Hamiltonian $H(g_1,g_2...g_n)$ such that $\mathbb{T}H(g_1,...g_n)=H(\tilde{g}_2...\tilde{g}_n)$ and on couplings $\mathbb{T}\lbrace g_1,g_2,...g_n\rbrace=\lbrace \tilde{g}_1,\tilde{g}_2,...\tilde{g}_n\rbrace$ where $\forall g_i \in \mathbb{C}$ mapped Hamiltonian does not obey the c and g theorems in general.\\

$\mathbb{T}^{\alpha}_{n}\lbrace \tilde{\bold{g}} +\delta\bold{ g}\rbrace=\tilde{\bold{g}}+\sum_{n}\delta\bold{g}_{n}\tilde{O}(\alpha)_n + \sum_{n,m}\delta\bold{g}_{n}\tilde{O}(\alpha)_n\delta\bold{g}_{m}\tilde{O}(\alpha)_m +... $\\
$\tilde{O}$ is the eigenvector at the fixed point of the hamiltonian H. If we now use the bi-Orthonormal vectors for the non-Hermitian system pose the real eigenvalues then there exist a unitary operator $\theta$ as the following,\\
\begin{equation}
    \begin{split}
        \theta^\dag\theta\mathbb{T}^{\alpha}_{n}\theta^\dag\theta\lbrace \tilde{\bold{g}} +\delta\bold{ g}\rbrace&=\tilde{\bold{g}}+\sum_{n}\theta^\dag\theta\delta\bold{g}_{n}\theta^\dag\theta\tilde{O}(\alpha)_n +\\ &\sum_{n,m}\theta^\dag\theta\delta\bold{g}_{n}\theta^\dag\theta\tilde{O}(\alpha)_n\theta^\dag\theta\delta\bold{g}_{m}\theta^\dag\theta\tilde{O}(\alpha)_m +... 
    \end{split}
\end{equation}
The unitary rotation of the vector is $\theta \tilde{O}=\tilde{a}\tilde{O}_L+\tilde{b}\tilde{O}_R$ which ensures $\langle \tilde{O}_\eta|\tilde{O}_{\eta'}\rangle=\delta_{\eta\eta'}$ and the coupling matrix is $\bold{g}=\begin{pmatrix}g_{real}\\g_{im}\end{pmatrix}$ and vector $O_n$ is at fixed point $\tilde{O}_n=\begin{pmatrix}
     \tilde{O}_L\\\tilde{O}_R\end{pmatrix}$  This also ensure the under RG flow if c and g theorems violated then we have $\frac{\delta \tilde{O}_{L}}{\delta \tilde{O}_R}\to \frac{\delta g_{real}}{\delta g_{im}}\to$ log spiral along the invarient (from above corollary 1.4) any deviation to this the information not conserved due to the loss of invarience.The direction of the spiral depends on the RG transformation on vector it may be either inward or outward.
\subsection{Complex Closed Orbits in Quantum Mechanics}
We discuss some fundamental theorems in basic mathematics what are the scenario generically to get the limit cycle in complex plane. We explore If the wave function having the closed trajectory in complex plane does it guaranty the integrability. One can propose the following  condition along with norm monotone  for getting the closed orbit.\\
\textbf{\textit{Proposition 1}} A closed trajectory namely $f(z-a)\overline{f(z-a)}-a=0$ in complex plane is possible  iff $f(z-a)$, $\forall a \in \mathbb{R}$ is analytic and has a Green theorem on vector bundle $V(x,y)$ $\in$ $\mathbb{C}$ as $\oint \frac{\big(f(z)\frac{\partial v}{\partial y}-f(\bar{z})\frac{\partial v}{\partial x}\big)-i\big(f(z)\frac{\partial u}{\partial y}+f(\bar{z})\frac{\partial u}{\partial x}\big)}{V(x,y)}d\tau=0$ where $\bar{f}=f(\bar{z})$ and $V(x,y)=u(x,y)+iv(x,y)\neq 0$.\\
For real partial differential equations the Bendixson ,Bendixson-Dulac theorems are proved by contradiction and by integrand vanishing criterian for fixed sign in the $\oint \frac{d\log V}{d\tau}d\tau=-\oint (\frac{\partial f(z)}{\partial x}+\frac{\partial f(\bar{z})}{\partial y})d\tau$   for real solutions\cite{gine2014non}. It becomes challenging to show right hand side has the fixed sign in the complex plane.\\
Given the above proposition, if the closed orbit exists in the complex plane, then we always have $\oint \psi^*\psi d\tau <\infty $ square integrability and a vector space V . The converse need not be true always.

\section{Analytic solution of complex oscillator}
Let's consider a standard Harmonic Oscillator( all the constants set to unity except complex interaction) with a complex interaction which depends on position operator as an example. Although we discuss very general methods can be applied everywhere.,
\begin{equation}
\label{eq:model}
    \begin{split}
        \bigg(-\frac{d^2}{dx^2}+x^2+(i\gamma)^{2N} x^{2N}\bigg)\psi=E\psi
    \end{split}
\end{equation}
If the above differential equation posses a solution then it must have the analytic properties then we can choose a solution $\psi=e^{\frac{-x^{2N+2}}{(2N+1)(2N+2)}}\alpha(x)$ as a ansatz which can be derived by taking large x limit of eq \ref{eq:model}.With bit of algebra we get the following,

\begin{equation}
    \begin{split}
        \alpha''+2\frac{x^{2N+1}}{2N+1}\alpha'-\bigg(x^2-\frac{x^{4N+2}}{(2N+1)^2}+(i\gamma)^{2N}x^{2N}-E\bigg)\alpha=0
    \end{split}
\end{equation}
This above reduces to Hermite differential equation for $N\to 0$ limit. This is a complex ODE,so the solution consist both real and imaginary parts we can use a Frobenius method as $\alpha=\sum^{\infty}_{n=0}c_n x^n e^{in\theta(x)}$ this will yield the following,
\begin{equation}
\label{eq:pro}
    \begin{split}
    \sum_{n}c_{n}n(n-1)x^{n-2}e^{in\theta(x)}+\sum_{n}2\frac{x^{2N+1}}{2N+1}nc_nx^{n-1}e^{in\theta(x)}\\
    -\sum_{n}\bigg(x^2-\frac{x^{4N+2}}{(2N+1)^2}+(i\gamma)^{2N}x^{2N}-E\bigg)c_nx^n e^{in\theta(x)}=0\\
    n^2\frac{d^2\theta}{dx^2}+in\frac{x^{2N+1}}{2N+1}\frac{d\theta}{dx}-\beta(x)=0\\
    where\hspace{2mm} \beta(x)=\bigg(x^2-\frac{x^{4N+2}}{(2N+1)^2}+(i\gamma)^{2N}x^{2N}-E\bigg)
    \end{split}
\end{equation}
\textbf{\textit{validity of the frobenius method}} can be discussed separately there are various lemma associated to the complex disc and analytic continuation along the singular points, although the convergence does not guaranty but still we can use the method, for instance if we do a convergence test all the terms in the recurence relation vanish except $\lim_{N\to \infty}\bigg|\frac{c_{n+2N}}{c_{n-2N}}\bigg|=\lim_{N\to\infty}\bigg|\frac{n-2N}{(i\gamma)^{2N}}e^{-2i\theta}\bigg|$ as discussed in the appendix. This does not affect the solution since this is entire function in complex domain.
For $\theta$ differential equation we can find the integrating factor,
\begin{equation}
\label{eq:theta}
    \begin{split}
        \theta'e^{\frac{x^{2N+1}}{2N+1}}=\int\beta(x)e^{\frac{x^{2N+2}}{(2N+1)(2N+2)}}dx +c
    \end{split}
\end{equation}
This solution \ref{eq:theta} gives nice forms which will be important in later analysis,
\begin{equation}
    \begin{split}
        \int\beta(x)e^{\frac{x^{2N+2}}{(2N+1)(2N+2)}}dx&=\int x^2e^{\frac{x^{2N+2}}{(2N+1)(2N+2)}}dx-\int\frac{x^{2N+2}}{(2N+1)^2}e^{\frac{x^{2N+2}}{(2N+1)(2N+2)}}dx\\
        &+\int (i\gamma)^{2N}x^{2N}e^{\frac{x^{2N+2}}{(2N+1)(2N+2)}}dx
        -\int Ee^{\frac{x^{2N+2}}{(2N+1)(2N+2)}}dx
    \end{split}
\end{equation}
The integral on the right hand side  can be transformed  to the following,
\begin{equation}
    \begin{split}
        &I=k\bigg(\int (at)^{\frac{1-2N}{2N+1}}e^{t}dt-\int\frac{(at)^{\frac{2N+1}{2N+2}}}{(2N+1)^2}e^{t}dt
        +c\int (i\gamma)^{2N}(at)^{\frac{-2}{2N+2}}e^{t}dx-d\int E(at)^{-\frac{2N+1}{2N+2}}e^{t}dt\bigg)\\
        &where \hspace{2mm} a=(2N+1)(2N+2), k=2N+1,t= \frac{x^{2N+2}}{(2N+1)(2N+2)}
    \end{split}
\end{equation}
These can be expressed as incomplete Gamma function
\begin{equation}
    \begin{split}
        I=k\bigg(-E (-a)^{-c} \Gamma_{-c,t} +(i \gamma )^{2 N} (-a)^{-d} \Gamma_{d,t}
        +(-a)^{b} \Gamma_{b,t}-\frac{(-1)^{-c} a^{c} \Gamma_{c,t}}{(2 N+1)^2}\bigg)\\
        \Gamma_{-c,t}= \Gamma \big(1-c,-t\big),\Gamma_{d,t}=\Gamma \big(1-d,-t\big),\Gamma_{b,t}=\Gamma \big(b+1,-t\big),\Gamma_{c,t}=\Gamma \big(c+1,-t\big)\\
        where \hspace{2mm} b=\frac{1-2N}{2N+2},c=\frac{2N+1}{2N+2},d=\frac{1}{N+1}
    \end{split}
\end{equation}

Complete  solution of the $\theta$ function  can be derived as the following,
\begin{equation}
\label{eq:solntheta}
    \begin{split}
        \theta=-\frac{  (a t)^{c} \left(-2(N+1)t+(4 N+3)e^{-t} \Gamma_1\right)}{(2 N+1) (4 N+3)}
        -E k  (a t)^{-c} \left(-2(N+1)t+e^{-t}t^{c}\Gamma_2\right)\\
        -\frac{1}{2} k (a t)^{b} \left(-(2 N+1)t+e^{-t}(-t)^{-b} \Gamma_3\right)
        -\frac{k  (i \gamma )^{2 N}  (a t)^{-d} \left(-(N+1)t+e^{-t}t^dN \Gamma_4\right)}{N}\\
        \Gamma_1=\Gamma \left(\frac{4 N+3}{2 N+2},-t\right),\Gamma_2=\Gamma \left(\frac{1}{2 N+2},-t\right),\\\Gamma_3=\Gamma \left(\frac{2}{2 N+1},-t\right),\Gamma_4=\Gamma \left(\frac{N}{N+1},-t\right)
    \end{split}
\end{equation}
The theta solution becomes important in renormalization context as we know action yields the phase information of the wave function which is discussed in wave function renormalization. Rewriting the full wave function we have the following,
\begin{equation}
\label{eq:wavefn}
    \begin{split}
        \psi=e^{-\frac{x^{2N+2}}{(2N+1)(2N+2)}}\sum_{n}c_{n}x^{n}\prod^{4}_{i=1}e^{in\tilde{\Gamma}_i}\prod^{4}_{j=1}e^{inf_{j}(x)}
    \end{split}
\end{equation}
In eq \ref{eq:wavefn} the $\Gamma_i$ represent the incomplete gamma functions (as in eq \ref{eq:solntheta}) derived earlier and $f_j(x)$ are the algebraic functions(as in eq \ref{eq:solntheta}) generally polynomials in t where t in this section defined earlier.
As we know the condition for coefficients from eq\ref{eq:pro} the Frobenius solution for given N, we can find coefficients.
\section{Unitary and Non-Unitary Regimes}
It is necessary to investigate whether the Unitary regime exist throughout the all time limits. If we preserve the symmetry in all parameter regimes then is there a possibility still we go into non-unitary regime. We start with the quantum master equation to address this.
\begin{equation}
    \begin{split}
        \frac{d\hat{\rho}}{dt}=-i[H,\hat{\rho}]+\kappa(2\mathcal{L}\rho\mathcal{L}^{\dag}-\lbrace\mathcal{L}^{\dag}\mathcal{L},\rho\rbrace)
    \end{split}
\end{equation}
We can see the above density matrix can have a unitary dynamics when the recycling and other term vanishes,
\begin{equation}
\label{eq:cond}
    \begin{split}
        2\mathcal{L}\rho\mathcal{L}^\dag-\lbrace\mathcal{L}\mathcal{L}^\dag,\rho\rbrace=0
    \end{split}
\end{equation}
we can construct a density matrix and Lindbladians as the following,
\begin{equation}
\label{eq:lind}
    \begin{split}
        \mathcal{L}=i\frac{\partial}{\partial x} -ie^{i\theta}\frac{   \partial}{\partial \theta}+e^{i\theta}x\\
        \rho=|x\rangle\langle x'|\otimes |\theta\rangle\langle\theta'|
    \end{split}
\end{equation}
 If we compute the eq \ref{eq:cond} by substituting the eq \ref{eq:lind}
 \begin{equation}
     \begin{split}
        2\mathcal{L}\rho\mathcal{L}^\dag= 2e^{-i\theta}\frac{\partial}{\partial x}\bigg(e^{-i\theta}\frac{\partial \rho_{\theta}}{\partial x} \otimes \rho_{x} +e^{-i\theta}\rho_{\theta}\otimes\frac{\partial \rho_{x}}{\partial x} \bigg)
         -2ie^{-i\theta}\frac{\partial}{\partial \theta}\bigg(e^{-i\theta}\frac{\partial \rho_{\theta}}{\partial \theta} \otimes \rho_{x} -e^{-i\theta}\rho_{\theta}\otimes \rho_{x} \bigg)\\
         +2e^{-i\theta}\frac{\partial}{\partial x}\bigg(e^{-i\theta}\frac{\partial \rho_{\theta}}{\partial \theta} \otimes \rho_{x} -e^{-i\theta}\rho_{\theta}\otimes \rho_{x} \bigg)
         -2ie^{-i\theta}\frac{\partial}{\partial \theta}\bigg(e^{-i\theta}\frac{\partial \rho_{\theta}}{\partial x} \otimes \rho_{x} +e^{-i\theta}\rho_{\theta}\otimes\frac{\partial \rho_{x}}{\partial x} \bigg)
     \end{split}
 \end{equation}
 Above set of equations gives the condition that these terms will vanish when $\rho_{\theta}\otimes\rho_{x}-\rho_{x}\otimes\rho_{\theta}=0$ and $(\theta^{-1}\log\rho_{\theta}-\mathbb{I})\otimes \rho_{x}=0$ this suggest the solution for $\rho_{\theta}=e^{c_1 \theta+c^{*}_1 \theta'}|0\rangle\langle0|$.The
 second term in the eq \ref{eq:cond} gives the following,
 \begin{equation}
     \begin{split}
         \bigg(e^{i\theta}\frac{\partial^2}{\partial x^2}e^{-i\theta}+e^{i\theta}\frac{\partial^2}{\partial \theta^2}e^{-i\theta}-ix\frac{\partial\theta}{\partial x}+(|c_1|^2-x^2)\bigg)\rho=0
     \end{split}
 \end{equation}
 This gives the following since the phase is also pure function of the position it follows $[x,f(\theta)]=0$ the phase operator equation reduces to the following,
 \begin{equation}
     \begin{split}
         \frac{\partial^2 \theta}{\partial x^2}+ix\frac{\partial \theta}{\partial x}+(|c_1|^2-x^2)=0
     \end{split}
 \end{equation}
 This above phase equation just of the form in eq \ref{eq:pro} except the $\beta(x)$ is different and this also satisfies the incomplete gamma functions as a solution.These incomplete gamma functions can be expressed in the hyper-geometric functions\cite{ozarslan2019some}. The $\theta$ complex ODE can be generalized as complex Hermite equation and we can get visualize the closed trajectories in the complex plane. 
 \begin{figure}
    \centering
    \includegraphics[scale=0.7]{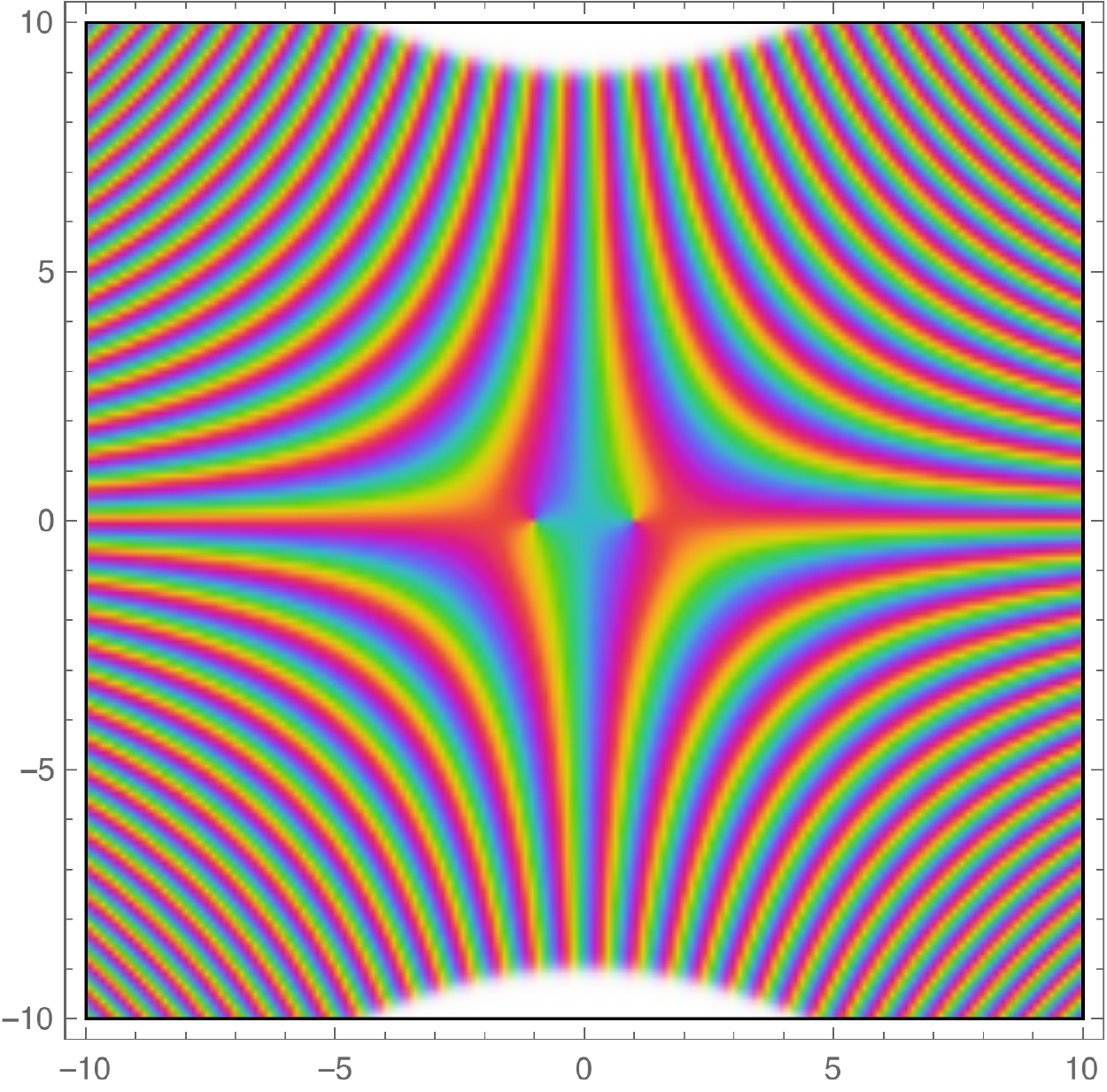}
     \caption{Visualization of the theta function in the complex plane unitary condition one would always expect such a phase function in the unitary regime. This we will also confirm from conventional RG calculation on action gives only the unitary regime, Hence the non-unitary regimes will not be captured by conventional methods. $\theta \propto \Gamma \left(\frac{N+1}{2},-\frac{1}{2} \left(i x^2\right)\right)$ the significant deviation from this appear for large N. }
    \label{fig:visual}
\end{figure}
 \begin{figure}
    \centering
    \includegraphics[scale=0.70]{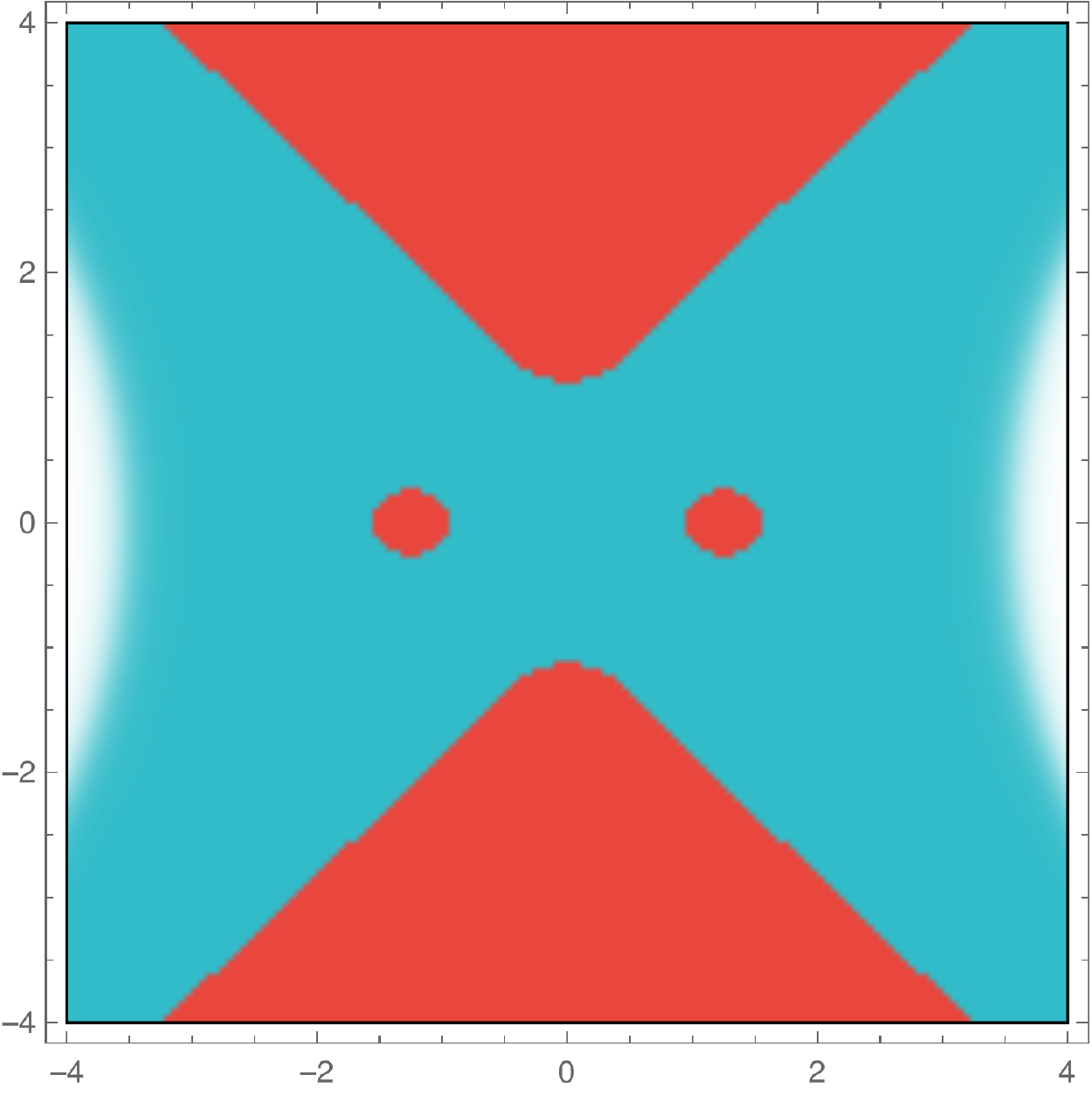}
     \caption{Visualization of the $H(z)H^{*}(z)-k$ for the theta function which are solutions of complex Hermite ODE, This turns out very crucial orthogonal polynomial(square integrable) in complex plane can give raise to closed trajectories.}
    \label{fig:visual}
\end{figure}
\subsection{Time evolution of the density matrix}
We now evaluate the time evolution from the possible density matrix we derived from Unitary condition and density matrix is $e^{c_1 \theta+c^{*}_1\theta'}\rho_{x}\otimes|0\rangle\langle0|$
\begin{equation}
    \begin{split}
        \frac{\partial \rho_x}{\partial t}\otimes\rho_{\theta}+\frac{\partial c_1}{\partial t}\rho_{x} \otimes \rho_{\theta}+\frac{\partial c^{*}_{1}}{\partial t}\rho_{x}\otimes\rho_{\theta}+ic_{1} \frac{\partial \theta}{\partial t}\otimes\rho_{\theta}\\
        ic^{*}_{1} \frac{\partial \theta}{\partial t}\otimes\rho_{\theta}=[H,\rho_x]\otimes\rho_{\theta}+\rho_x \otimes[H,\rho_\theta]=k\rho_x\otimes\rho_\theta
    \end{split}
\end{equation}
we can separate these tensor ODE in following fashion,
\begin{equation}
    \begin{split}
        \frac{\partial\rho_x}{\partial t}\otimes \mathbb{I}=k\rho_x\otimes\mathbb{I},\hspace{2mm}
        \frac{\partial c_1}{\partial t}-k_1c^{*}_{1}-k=0\\
         \frac{\partial c^{*}_1}{\partial t}-k_1c^*_{1}=0,\hspace{2mm} \frac{\partial \theta}{\partial t}-k_1\theta=0
    \end{split}
\end{equation}
These above time evolution for a commuting and scalable position operator respect the solution $c^{*}_{1}=e^{k_1t}c^*_{1}(0) ,c_{1}=kte^{-k_1t}c_1 (0), \rho_{x}=e^{-kt}\rho_{x}(0),\theta(t)=-ikt+\theta_{0}$ this will suggest the density matrix evolution as the following,
\begin{equation}
    \begin{split}
        \rho(t,t')=e^{-i|k|(t- t')} e^{ \xi(t)kte^{-k_1t}c_1+\xi^{*}(t')e^{k^*_1t}c^*_1}\rho_{x}\otimes\rho_\theta\\
        \xi(t)=-ikt+\theta_0,\xi^{*}(t')=ikt'+\theta'_0
    \end{split}
\end{equation}
for small integral constant $k_1\to 0$ above time evolution can be Fourier transformed and can be showed the deviations from Unitary regimes(conventional delta and Lorentzian),
\begin{equation}
    \begin{split}
      \rho(\omega) \propto \text{bei}_{-\frac{1}{3}}\left(\frac{2 \left| \omega \right| ^{3/2}}{3 \sqrt{3} \sqrt{i k^2}}\right) +J_{-\frac{1}{3}}\left(\frac{2 \left| \omega \right| ^{3/2}}{3 \sqrt{3} \sqrt[4]{k^4}}\right)
      +\, _1F_4\left(1;\frac{7}{6},\frac{4}{3},\frac{5}{3},\frac{11}{6};\frac{\omega ^6}{2.18^3 k^4}\right)
    \end{split}
\end{equation}
These above functions are special functions such as Kelvin real,imaginary functions,Bessel and Hyper geometric functions both regularized and unregulated. This shows these theta functions when they deviate from unitary regime we do get these special solutions these appear at the lowest loop RG calculations later. 
We can show that any complex extension of a quantum mechanical model may or may not have the unitary regimes and often convectional methods does not capture this. Now we are set to do functional analysis and this complex problem can be renormalized and get the ground state properties.
\section{Bethe Ansatz for complex Oscillator and Gross Pitaevskii(GP) equation}
Bethe Ansatz for the quasi momentum(which is decomposed as $p=p_x +p_\theta$ $\therefore p=\frac{1}{i}\frac{\psi '}{\psi} +\frac{\theta'}{\theta}$ the prime correspond to the derivative with x) can be derived as the following,
\begin{equation}
\label{eq:bethe}
    \begin{split}
        p^2_{\theta}-i p'_{\theta}=-2\sum^{n}_{j=1} i^{2N}(x-x_j)^{2N}\\
        for \hspace{2mm} x\to \infty \hspace{2mm} p_x=ix+\frac{1}{i}\sum^{N}_{k=1}\frac{1}{x-x_k}
    \end{split}
\end{equation}
The above $\theta$ part of the quasi-momentum ansatz is of the Riccati equation which can be reduced to the second order ODE to solve for arbitrary function. substituting the $p_\theta=iu'/u$ will give following Riccati equation,
\begin{equation}
    \begin{split}
        u''-(i)^{2N}\sum^{n}_{j=1} (x-x_j)^{2N}u=0
    \end{split}
\end{equation}
This will give the solution in terms of modified Bessel first and second kind,
\begin{equation}
    \begin{split}
        u=\begin{cases}
a\sqrt{x}\mathcal{J}_{\frac{1}{2q}}\bigg(\frac{\sqrt{-\zeta}}{q}x^q\bigg)+a\sqrt{x}\mathcal{Y}_{\frac{1}{2q}}\bigg(\frac{\sqrt{-\zeta}}{q}x^q\bigg), &  \zeta >0,\\
a\sqrt{x}\mathcal{I}_{\frac{1}{2q}}\bigg(\frac{\sqrt{\zeta}}{q}x^q\bigg)+a\sqrt{x}\mathcal{K}_{\frac{1}{2q}}\bigg(\frac{\sqrt{\zeta}}{q}x^q\bigg), &  \zeta <0, 
\end{cases}
    \end{split}
\end{equation}
Where in above $q=\frac{N+2}{2}$, a is integral constant and $\zeta = i^{2N}$. This gives the ansatz for the $p_\theta$ as 
\begin{equation}
    \begin{split}
        p_\theta=i\frac{\sum^n_{j=1}u'(x-x_j)}{\sum^n_{j=1}u(x-x_j)}
    \end{split}
\end{equation}
Using the reccurence relation and derivatives for the bessel functions namely,
\begin{equation}
\label{eq:reccur}
    \begin{split}
        \mathcal{J}'_\nu=\frac{\nu}{z}\mathcal{J}_{\nu}-\mathcal{J}_{\nu+1}=\mathcal{J}_{\nu-1}-\frac{\nu}{z}\mathcal{J}_{\nu}\\
         \mathcal{I}'_\nu=\frac{\nu}{z}\mathcal{I}_{\nu}+\mathcal{I}_{\nu+1}=\mathcal{I}_{\nu-1}-\frac{\nu}{z}\mathcal{I}_{\nu}\\
         \mathcal{K}'_\nu=\frac{\nu}{z}\mathcal{K}_{\nu}-\mathcal{K}_{\nu+1}=-\mathcal{K}_{\nu-1}-\frac{\nu}{z}\mathcal{K}_{\nu}\\
         where \hspace{2mm} '\to \frac{\partial}{\partial z},\hspace{2mm} z\to x^q \implies \frac{\partial}{\partial z}=\frac{\partial x}{\partial z}\frac{\partial}{\partial x}
    \end{split}
\end{equation}
In our case the argument has some power($z=x^q$) hence above relations will be used with a scaling function $f(x)=qx^{q-1}$. Bessel $\mathcal{J}$ and $\mathcal{Y}$ have the same relations.
From the Ansatz we can reconstruct the wave function as the following given the bethe roots,
\begin{equation}
    \begin{split}
        \psi=e^{-\frac{x^2}{2}}\prod^{n}_{j=1}(x-x_j)\exp\bigg((i)^{2N}\frac{(x-x_j)^{2N+1}}{2N+1}\bigg)\\
        x_j=\frac{1}{2}\sum_{k\neq j}\frac{1}{(x_j-x_k)}+i^{2N}\sum_{k\neq j}(x_j-x_k)^{2N}
    \end{split}
\end{equation}
The first few roots can be computed analytically and show that the complex Hermite polynomials as the solution, but if we go beyond $2N > 2$, the solution for excited states becomes trickier also the validity of the ansatz and integrability of the problem need to be reconsidered.

\subsection{Connection to GP equation}
\begin{figure}
    \centering
    \includegraphics[scale=.75]{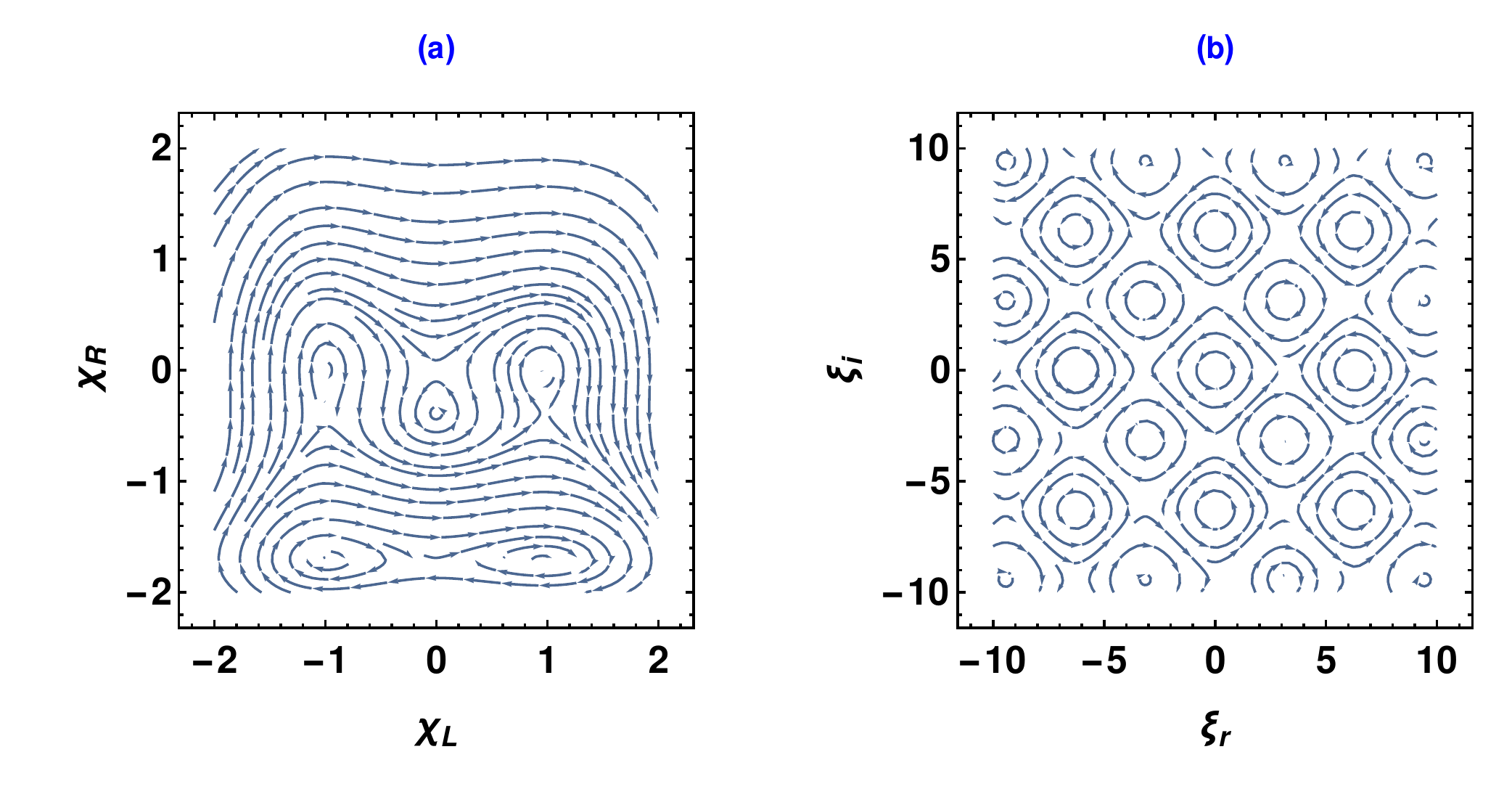}
    \caption{We can see even at the lowest power $2q=1$ there are limit cycles (both in space and time) in the Riccati ODE or the complex GP equation.As the roots grow this will destroy the limit cycles and vortexes disappear for larger complex strength($im(s')\to \infty$) asymptotically one vortex remain at origin.}
    \label{fig:GP_RG}
\end{figure}

The Riccati equation and its analytic solution has nice connection to the GP equation as discussed in the article\cite{al2010special}. The Bethe ansatz at large x the $p_x$ part will give the simple form but the  $p_\theta$ will retain the Riccati ODE form (This can be derived from time dependent Schrodinger equation and substituting the Bethe ansatz in eq \ref{eq:bethe}),
\begin{equation}
\label{eq:GP}
    \begin{split}
        \frac{\dot{u}}{1- \epsilon \dot{u'}/u}+ u''-(i)^{2N}\sum^{n}_{j=1} (x-x_j)^{2N}u =0\\
        as \hspace{2mm} \epsilon \to 0 \hspace{2mm} \dot{u'}/u=\frac{1}{u}\frac{\partial^2 u}{\partial x \partial t} \to 0
    \end{split}
\end{equation}
The above equation mimics complex GP in the specified limit though the time interval for this can not be specified as the usual GP, but one can note as the complex potential vanishes GP form will not be intact.\\
Keeping the Riccati form, what can we say about the complex GP equation is something very interesting. If we Renormalize the \ref{eq:GP} by separating the solutions in two limits as the following,
\begin{equation}
    \begin{split}
        u(x)&=u(x)|_{x<0}+u(x)|_{x>0}\\
        &=u_L+u_R
    \end{split}
\end{equation}
The solutions for \ref{eq:GP} can be written as $u=u(t)u(x)$ and to renrmalize them we can scale as $\tilde{x}\to sx$,$\tilde{t}\to s' t$,$\tilde{u(t)}\to \xi(s')u(t)$and $\tilde{u(x)}\to \chi(s)u(x)$. We derive the following RG equations for first few roots of the Bethe ansatz.We can even ignore the Bethe roots for the ground state and along with recurrence relations in eq \ref{eq:reccur} we get these eqns with complex flow parameters.
\begin{equation}
    \begin{split}
        \partial_s{\chi(s)}_{L,R}=s\chi_{L,R}(s)+s^{2q-1}\chi'_{L,R}(s)\\
        \partial_{s'}{\xi(s')}=\xi(s')
    \end{split}
\end{equation}
Where the $\chi'$ correspond to derivative with the s, We compute the RG trajectories for the $2q=1$ for simplicity to see the renormalization trajectories as shown in fig \ref{fig:GP_RG}.
\begin{figure}
    \centering
    \includegraphics[scale=1.0]{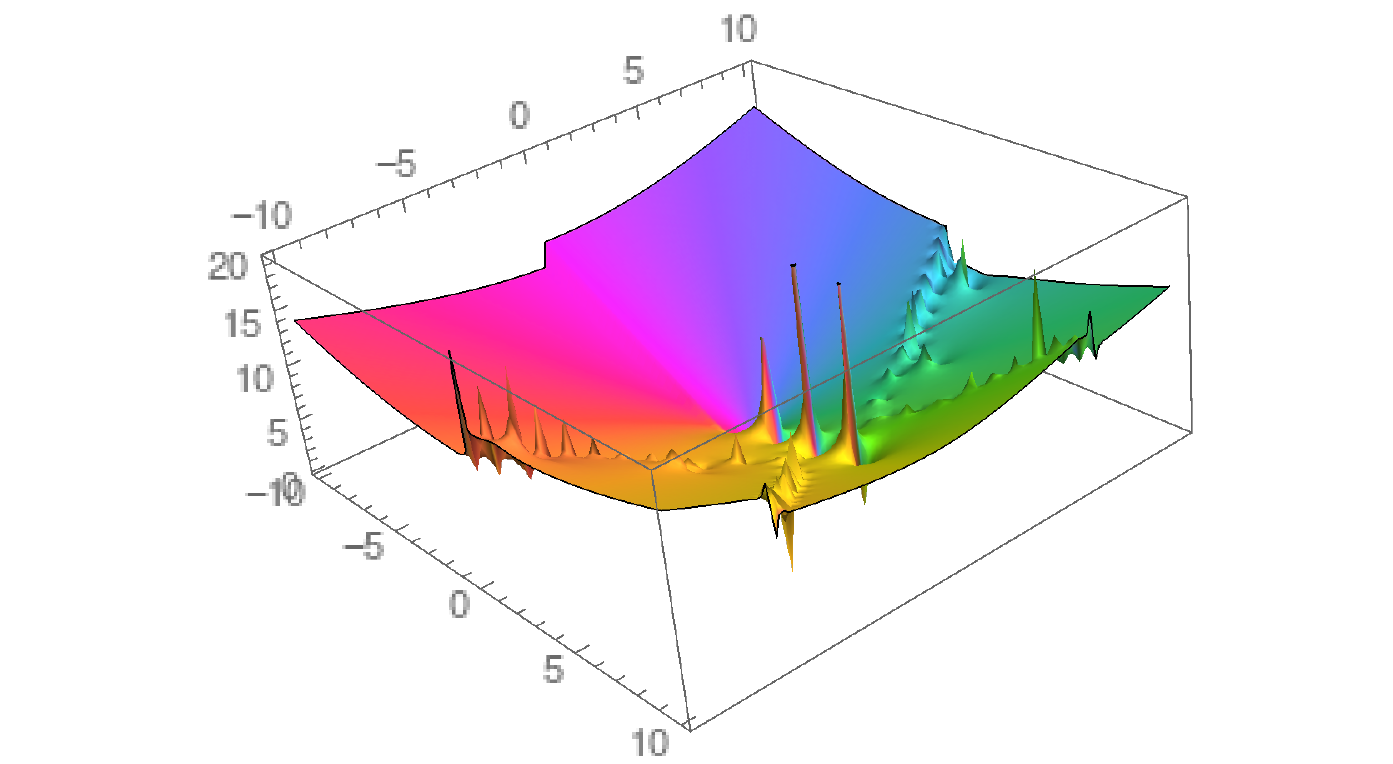}
    \caption{The real line poles in momentum ($p=p_x+p_\theta$)  ($n=3$ which are the eigen-enrgies) correspond to the spectrum of oscillator as complex strength increase there will be scattering states which mainly arise from the $p_\theta$ part}
    \label{fig:my_label}
\end{figure}
\section{FRG of quantum models} 
As we discussed earlier, the action RG should visualize generically phase$(\theta)$ of the wave function which are the  in the energy plane since we have real and imaginary parameters; both are functions of energy. We are expecting the trajectories of $\theta$ functions from our unitary conditions in complex energy surfaces.  No matter what regularization we use in usual methods, they do not capture the deviations from the Unitary regimes.
\subsection{Wave function Renormalliaztion}
It is very important to bring the connection between the action measure and the wavefunction.In earlier section briefly we discussed the RG beta functions connection to the eigenvectors.If we scale the $\tilde{\phi}=\sqrt{Z}\phi$ then the two point correlator will be $\langle\tilde{\phi}^\dag\tilde{\phi}\rangle=\frac{\delta^2 \mathcal{Z}}{\delta\tilde{\phi}^\dag \delta\tilde{\phi}}=\frac{1}{Z}G_{0}$ Scaling the left and right fields with the $\sqrt{Z_L}$ and $\sqrt{Z_R}$ with a unitary we can show the $\tilde{G} \to \frac{1}{\sqrt{Z^2_L+Z^2_R}}\begin{pmatrix}
     Z_{L}+Z_{R} & 0\\
     0& Z_{L}-Z_{R}
\end{pmatrix}G_0$ then we have,
\begin{equation}
    \begin{split}
        \frac{d Tr\tilde{G}}{d\Lambda}=0\implies
        \frac{\partial \det (e^{G})}{\partial \Lambda}e^{Tr(G)}\frac{\partial G}{\partial \Lambda}=0\\
        \frac{\partial Z_{R}}{\partial Z_{L}}=\frac{e^{Z_{R}+Z_{L}}}{e^{Z_{L}-Z_{R}}}\implies Z_{L}=e^{Z_R} 
    \end{split}
\end{equation}
The exact renormalization would capture the spiral as invarient provided that the RG procedure dont hit the determinant to zero. \\
\subsection{Equivalence of vertex and coordinate swapping operation}
The date,Jimbo,Kashiwara and Miwa (DJKM)\cite{jimbo1981transformation,gatto2020cohomology,gatto2021schubert,van2012ckp,behzad2020bosonic} Boson do exhibit bosonic Fock space polynomial algebra as $B(\zeta):=B\otimes\mathbb{Q}[\zeta^{-1},\zeta]$ .This is indeed the Lie super algebra and which is shown to be all basis elements correspond to finitely many exterior algebra or the Grassmann algebra.This allows us to expand the bosonic fock space as the elements of g-algebra.Also earlier works show the fermion-boson correspondence.The dual vector space for the bosons can be defined as $\mathcal{V}=\bigoplus_{i\in\mathbb{Z}}\mathbb{Q}.b_i$ and its dual will have the
$\mathcal{V}^*=\bigoplus_{i\in\mathbb{Z}}\mathbb{Q}.\beta_j$ where $\beta_{j}\in Hom_{\mathbb{Q}}(\mathcal{V},\mathbb{Q})$ this is homiomorphism of the $\mathbb{Q}$ -vector with the polynomial which has the linear form $\beta_j(b_i)=\delta_{i,j}$ this gives us the bosonic vertex as the following,
\begin{equation}
    gl(\mathcal{V})=\mathcal{V}\otimes\mathcal{V}^*=\bigoplus_{i,j\in \mathbb{Z}}\mathbb{Q}.b_i\otimes\beta_j
\end{equation}

\begin{theorem}
 The map of the $\mathcal{Z}(z,w)$ on a bosonic vertex operator $V(\zeta)$ given by
 \begin{equation}
     \mathcal{Z}(z,w)|_{V(\zeta^m)}=\frac{z^m}{w^m}\frac{1}{1-\frac{w}{z}}exp\bigg(\sum_{n\ge1} x_n(z^n-w^n)\bigg)exp\bigg(\sum_{n\ge1} \frac{(z^{-n}-w^{-n})}{n}\frac{\partial}{\partial x_n}\bigg)
 \end{equation}
 Where $\zeta=\frac{z}{w}$ and all $x_n$ are coordinates.This DJKM bosons formulae basically states that the m-order polynomial $gl(\mathcal{V})$ in bosonic fock space maps to same order polynomial in boson vertex. 
\end{theorem}
The above theorem is well celebrated and proved in various way including the recent works\cite{behzad2020bosonic}.We use the above language to state and prove the following theorem.
\begin{theorem}
 The map on the bosonic fock space can be equivalent to permuting any two successive coordinates which will be $\mathcal{Z}(z,w)\to P(x_1,x_2)\mathcal{Z}(z,w)\to P(z,w)\mathcal{Z}$
\end{theorem}
\begin{proof}
\begin{equation}
\begin{split}
     P(x_1,x_2) \mathcal{Z}(z,w)=P(x_1,x_2)exp\bigg(\sum_{n\ge1} x_n(z^n-w^n)\bigg)exp\bigg(\sum_{n\ge1} \frac{(z^{-n}-w^{-n})}{n}\frac{\partial}{\partial x_n}\bigg)\\
   =\mathcal{Z}_1exp\bigg(x_2(z-w)+x_1(z^2-w^2)+(z^{-1}-w^{-1})\frac{\partial}{\partial x_2}+\frac{(z^{-2}-w^{-2})}{2}\frac{\partial}{\partial x_1}\bigg)\\
   =\mathcal{Z}_1exp\bigg((z-w)[x_2+x_1(z+w)]+(z^{-1}-w^{-1})[\frac{\partial}{\partial x_2}+(z^{-1}+w^{-1})\frac{\partial}{\partial x_1}]\bigg)
\end{split}
\end{equation}
Now we can do a variable transform of any arbitrarily successive to new coordinates as $(x_{m-1},x_{m})\to(\alpha_{m-1},\alpha_m)$ as the following,
\begin{equation}
    \begin{split}
        \begin{pmatrix}
             1 & w^{m-1}\\
             1 & z^{m-1}
        \end{pmatrix}\begin{pmatrix}
             x_m\\
             x_{m-1}
        \end{pmatrix}=\begin{pmatrix}
             w^{m-1} & 1\\
             z^{m-1} & 1
        \end{pmatrix}\begin{pmatrix}
             \alpha_m\\
             \alpha_{m-1}
        \end{pmatrix}\\
        \begin{pmatrix}
             1 & w^{-(m-1)}\\
             1 & z^{-(m-1)}
        \end{pmatrix}\begin{pmatrix}
             \frac{\partial}{\partial x_m}\\
             \frac{\partial}{\partial x_{m-1}}
        \end{pmatrix}=\begin{pmatrix}
             w^{-(m-1)} & 1\\
             z^{-(m-1)} & 1
        \end{pmatrix}\begin{pmatrix}
             \frac{\partial}{\partial \alpha_m}\\
             \frac{\partial}{\partial \alpha_{m-1}}
        \end{pmatrix}
    \end{split}
\end{equation}
Using the above representations we can write the permutation again as
\begin{equation}
    \begin{split}
        P(x_1,x_2) \mathcal{Z}(z,w)=\mathcal{Z}(w,z)|_{x_2}\mathcal{Z}(w,z)|_{\alpha_1,\alpha_2}\mathcal{Z}(w,z)|_{x_{n>3}}
    \end{split}
\end{equation}
Now recursively if we do permutations of successive coordinates,
\begin{equation}
    \begin{split}
        P(x_1,x_2) P(x_2,x_3)...P(x_{m-1},x_m)\mathcal{Z}(z,w)=\mathcal{Z}(w,z)|_{x_2,x_3,..x_m}\mathcal{Z}(w,z)|_{\alpha_1,\alpha_2...\alpha_m}\mathcal{Z}(w,z)|_{x_{n>m}}
    \end{split}
\end{equation}
when m hits n then we have the following, Let's call unidirectional successive swapping as $\mathbb{P}$ 
\begin{equation}
\begin{split}
     \mathbb{P}\mathcal{Z}(z,w)&=\mathcal{Z}(w,z)|_{x_2,x_3,..x_m}\mathcal{Z}(z,w)|_{\lbrace\alpha_n\rbrace}\\
    &=\mathcal{Z}(w,z)|_{x_1}\mathcal{Z}(w,z)|_{\lbrace x_n\rbrace}\mathcal{Z}(z,w)|_{\lbrace\alpha_n\rbrace}
\end{split}
\end{equation}
Using the inverse representation we can show the vertex algebra and permutation gives the same map on the bosonic fock space
\end{proof}
Using the theorem we can show the equivalence between the vertex-RG and permuting methods.
\subsection{Equivalence in the scalar field Wetterich FRG and the matsuberra RG}
Starting from a partition functional $\mathcal{Z}=\int \mathcal{D}[\phi] e^{-S[\phi]+J.\phi}$ using Schwinger formulation and  Legendre  transform we can derive the Polchinski equation there by Wetterich equation which is very standard.
\begin{equation}
    \begin{split}
        \mathcal{Z}=\int \mathcal{D}[\phi] e^{-S[\phi]+J.\phi}\\
        W_k=\ln \mathcal{Z}_k[J]=\ln \int \mathcal{D}[\phi]e^{-S[\phi]+J.\phi-\Delta S_k}  \\
        \partial_k W_k=-\frac{1}{2}\langle\phi .\partial_k R_k. \phi\rangle=-\frac{1}{2}(\langle\phi.\phi\rangle_c+\phi.\phi).\partial_kR_k\\
        \langle\phi.\phi\rangle_c=W^{(2)}_k=\frac{\delta^2 W_k}{\delta^2 J}=\frac{\delta \phi}{\delta J},\hspace{2mm}
        \tilde{\Gamma}_k=\underset{J}{\textbf{sup}}(J.\phi-W_k[J])\\
        \Gamma_k=\tilde{\Gamma}_k-\Delta S_k, \hspace{2mm} J_k=\frac{\delta \tilde{\Gamma}_k}{\delta \phi}\implies
        \partial_{\ln k}\Gamma_k=-\frac{1}{2}Tr\bigg(\frac{\partial_{\ln k} R_k}{\Gamma^{(2)}+R_k}\bigg)
    \end{split}
\end{equation}
The above can be represented in diagrammatically to generic n point function or the FRG loop with the insertion of regulator vertex and summing over the intermediate or the loop index.
\tikzset{every picture/.style={line width=0.75pt}} 

\begin{tikzpicture}[x=0.75pt,y=0.75pt,yscale=-1,xscale=1]

\draw  [dash pattern={on 0.84pt off 2.51pt}]  (250.25,139.97) -- (350.05,140.42) ;
\draw  [fill={rgb, 255:red, 155; green, 155; blue, 155 }  ,fill opacity=1 ] (277.36,139.99) .. controls (277.36,138.59) and (278.49,137.46) .. (279.89,137.46) .. controls (281.29,137.46) and (282.43,138.59) .. (282.43,139.99) .. controls (282.43,141.39) and (281.29,142.52) .. (279.89,142.52) .. controls (278.49,142.52) and (277.36,141.39) .. (277.36,139.99) -- cycle ;
\draw    (300.59,115.65) -- (304.59,120.22) ;
\draw    (304.79,115.76) -- (300.19,119.72) ;
\draw  [fill={rgb, 255:red, 155; green, 155; blue, 155 }  ,fill opacity=1 ] (321.48,140.29) .. controls (321.48,138.72) and (322.75,137.44) .. (324.32,137.44) .. controls (325.89,137.44) and (327.16,138.72) .. (327.16,140.29) .. controls (327.16,141.86) and (325.89,143.13) .. (324.32,143.13) .. controls (322.75,143.13) and (321.48,141.86) .. (321.48,140.29) -- cycle ;
\draw   (299.98,117.73) .. controls (299.98,116.33) and (301.12,115.2) .. (302.52,115.2) .. controls (303.92,115.2) and (305.05,116.33) .. (305.05,117.73) .. controls (305.05,119.13) and (303.92,120.27) .. (302.52,120.27) .. controls (301.12,120.27) and (299.98,119.13) .. (299.98,117.73) -- cycle ;
\draw  [draw opacity=0] (280.31,137.11) .. controls (282.29,126.57) and (290.25,118.52) .. (299.98,117.73) -- (301.56,142.23) -- cycle ; \draw   (280.31,137.11) .. controls (282.29,126.57) and (290.25,118.52) .. (299.98,117.73) ;
\draw  [draw opacity=0] (305.22,117.9) .. controls (315.61,119.85) and (323.58,127.61) .. (324.56,137.16) -- (300.1,139.15) -- cycle ; \draw   (305.22,117.9) .. controls (315.61,119.85) and (323.58,127.61) .. (324.56,137.16) ;
\draw  [draw opacity=0] (324.36,143.56) .. controls (323.96,154.78) and (314.09,163.77) .. (301.97,163.77) .. controls (289.64,163.77) and (279.64,154.47) .. (279.57,142.97) -- (301.97,142.85) -- cycle ; \draw   (324.36,143.56) .. controls (323.96,154.78) and (314.09,163.77) .. (301.97,163.77) .. controls (289.64,163.77) and (279.64,154.47) .. (279.57,142.97) ;

\draw  [dash pattern={on 0.84pt off 2.51pt}]  (513.8,172.99) -- (603.78,172.71) ;
\draw  [draw opacity=0] (552.1,172.47) .. controls (536.43,171.47) and (524.02,158.45) .. (524.02,142.53) .. controls (524.02,127.17) and (535.56,114.51) .. (550.43,112.74) -- (554.02,142.53) -- cycle ; \draw   (552.1,172.47) .. controls (536.43,171.47) and (524.02,158.45) .. (524.02,142.53) .. controls (524.02,127.17) and (535.56,114.51) .. (550.43,112.74) ;
\draw  [draw opacity=0] (555.95,112.59) .. controls (571.62,113.58) and (584.02,126.6) .. (584.02,142.53) .. controls (584.02,157.88) and (572.49,170.54) .. (557.61,172.31) -- (554.02,142.53) -- cycle ; \draw   (555.95,112.59) .. controls (571.62,113.58) and (584.02,126.6) .. (584.02,142.53) .. controls (584.02,157.88) and (572.49,170.54) .. (557.61,172.31) ;
\draw   (550.43,112.74) .. controls (550.43,111.18) and (551.7,109.92) .. (553.25,109.92) .. controls (554.81,109.92) and (556.07,111.18) .. (556.07,112.74) .. controls (556.07,114.3) and (554.81,115.56) .. (553.25,115.56) .. controls (551.7,115.56) and (550.43,114.3) .. (550.43,112.74) -- cycle ;
\draw  [fill={rgb, 255:red, 155; green, 155; blue, 155 }  ,fill opacity=1 ] (552.1,172.47) .. controls (552.1,170.91) and (553.36,169.65) .. (554.92,169.65) .. controls (556.47,169.65) and (557.73,170.91) .. (557.73,172.47) .. controls (557.73,174.02) and (556.47,175.28) .. (554.92,175.28) .. controls (553.36,175.28) and (552.1,174.02) .. (552.1,172.47) -- cycle ;
\draw    (551.47,110.84) -- (555.24,114.6) ;
\draw    (555.12,110.84) -- (551.35,114.6) ;

\draw  [dash pattern={on 0.84pt off 2.51pt}]  (485.71,140.62) -- (385.91,140.17) ;
\draw  [fill={rgb, 255:red, 155; green, 155; blue, 155 }  ,fill opacity=1 ] (458.6,140.59) .. controls (458.6,141.99) and (457.47,143.13) .. (456.07,143.13) .. controls (454.67,143.13) and (453.53,141.99) .. (453.53,140.59) .. controls (453.53,139.2) and (454.67,138.06) .. (456.07,138.06) .. controls (457.47,138.06) and (458.6,139.2) .. (458.6,140.59) -- cycle ;
\draw    (435.37,164.93) -- (431.37,160.36) ;
\draw    (431.17,164.83) -- (435.77,160.87) ;
\draw  [fill={rgb, 255:red, 155; green, 155; blue, 155 }  ,fill opacity=1 ] (414.48,140.3) .. controls (414.48,141.87) and (413.21,143.14) .. (411.64,143.14) .. controls (410.07,143.14) and (408.8,141.87) .. (408.8,140.3) .. controls (408.8,138.73) and (410.07,137.46) .. (411.64,137.46) .. controls (413.21,137.46) and (414.48,138.73) .. (414.48,140.3) -- cycle ;
\draw   (435.98,162.85) .. controls (435.98,164.25) and (434.84,165.38) .. (433.44,165.38) .. controls (432.04,165.38) and (430.91,164.25) .. (430.91,162.85) .. controls (430.91,161.45) and (432.04,160.32) .. (433.44,160.32) .. controls (434.84,160.32) and (435.98,161.45) .. (435.98,162.85) -- cycle ;
\draw  [draw opacity=0] (455.65,143.47) .. controls (453.67,154.02) and (445.71,162.06) .. (435.98,162.85) -- (434.4,138.36) -- cycle ; \draw   (455.65,143.47) .. controls (453.67,154.02) and (445.71,162.06) .. (435.98,162.85) ;
\draw  [draw opacity=0] (430.74,162.69) .. controls (420.35,160.74) and (412.38,152.97) .. (411.4,143.43) -- (435.86,141.44) -- cycle ; \draw   (430.74,162.69) .. controls (420.35,160.74) and (412.38,152.97) .. (411.4,143.43) ;
\draw  [draw opacity=0] (411.6,137.03) .. controls (412,125.8) and (421.87,116.81) .. (433.99,116.81) .. controls (446.32,116.81) and (456.32,126.12) .. (456.39,137.62) -- (433.99,137.74) -- cycle ; \draw   (411.6,137.03) .. controls (412,125.8) and (421.87,116.81) .. (433.99,116.81) .. controls (446.32,116.81) and (456.32,126.12) .. (456.39,137.62) ;

\draw (177.93,128.83) node [anchor=north west][inner sep=0.75pt]    {$\partial _{k} \Gamma _{k}^{( n)} =$};
\draw (354.04,129.55) node [anchor=north west][inner sep=0.75pt]    {$+$};
\draw (505.15,130.48) node [anchor=north west][inner sep=0.75pt]    {$-$};
\draw (286.99,97.58) node [anchor=north west][inner sep=0.75pt]    {$\partial _{k} R_{k}$};
\draw (416.85,165.84) node [anchor=north west][inner sep=0.75pt]    {$\partial _{k} R_{k}$};
\draw (244.07,118.64) node [anchor=north west][inner sep=0.75pt]    {$\Gamma _{k}^{( n+1)}$};
\draw (324.38,119.55) node [anchor=north west][inner sep=0.75pt]    {$\Gamma _{k}^{( n+1)}$};
\draw (372.4,118.02) node [anchor=north west][inner sep=0.75pt]    {$\Gamma _{k}^{( n+1)}$};
\draw (535.53,93.65) node [anchor=north west][inner sep=0.75pt]    {$\partial _{k} R_{k}$};
\draw (456.12,118.73) node [anchor=north west][inner sep=0.75pt]    {$\Gamma _{k}^{( n+1)}$};
\draw (541.81,176.28) node [anchor=north west][inner sep=0.75pt]    {$\Gamma _{k}^{( n+2)}$};

\end{tikzpicture}\\
For a special case $n=2$ of general diagrams represented above will generate the higher point correlation function as the following. 
\begin{equation}
\begin{split}
    \partial_{k}\Gamma^{(2)}=\frac{1}{2}Tr\bigg(\frac{\partial_k R_k}{\Gamma^{(2)}+R_k}\Gamma^{(3)}_k\frac{1}{\Gamma^{(2)}+R_k}\Gamma^{(3)}\frac{1}{\Gamma^{(2)}+R_k} \bigg)\\+\frac{1}{2}Tr\bigg(\frac{1}{\Gamma^{(2)}+R_k}\Gamma^{(3)}_k\frac{\partial_k R_k}{\Gamma^{(2)}+R_k}\Gamma^{(3)}\frac{1}{\Gamma^{(2)}+R_k} \bigg)\\-\frac{1}{2}
    Tr\bigg(\frac{1}{\Gamma^{(2)}+R_k}\Gamma^{(4)}_k\frac{1}{\Gamma^{(2)}+R_k} \bigg)
\end{split}
\end{equation}

\begin{figure}
    \centering
    \includegraphics[scale=1.2]{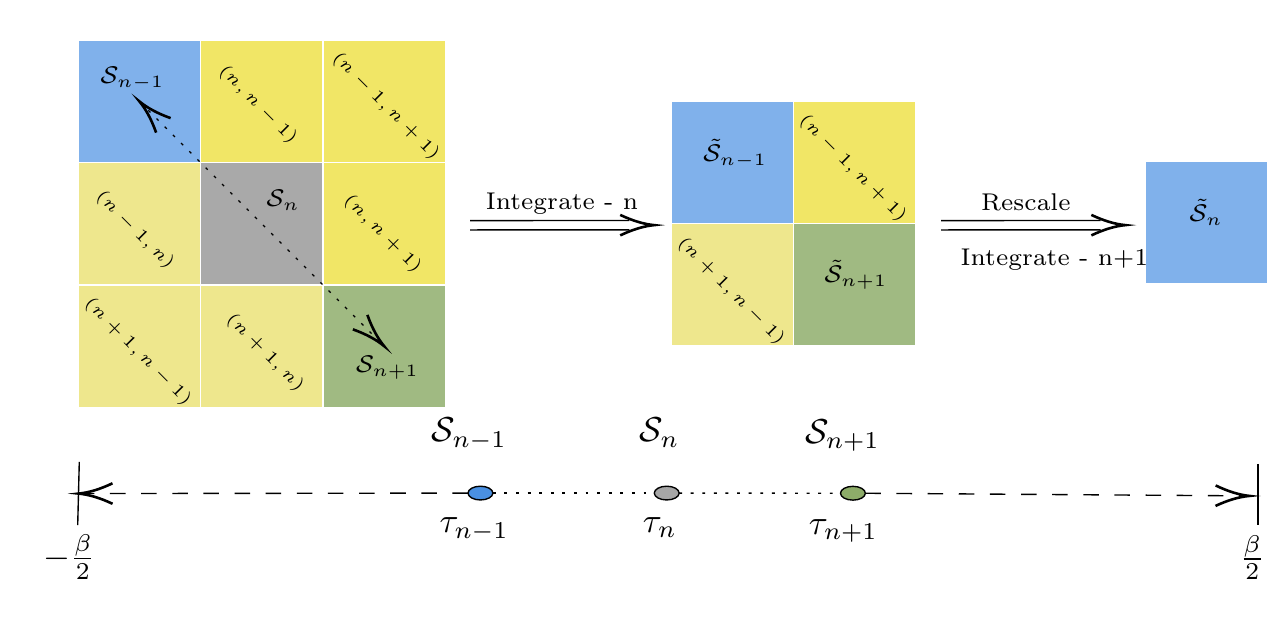}
     \caption{Schematic representation of the RG scheme we employed in $\tau$- space for Left-Right fields.}
    \label{fig:tau_RG}
\end{figure}

Exact flow equation for arbitrary n-point function  depends on the  (n+1) and (n+2) vertices  as represented above.We now show the similarity between the $\tau$-RG and  RG reccurence relations.We can start from the 
Bosonic Grassman which is the complex variable integral as the following,
\begin{equation}
    \begin{split}
        \mathcal{Z}=\int \prod_n \mathcal{D}[\bar{\phi}(\tau_n)\phi(\tau_n)]e^{\sum_{n}\bar{\phi}_l (\tau_n)(-\partial_{\tau_n}+\epsilon_l+R_k)\phi_l (\tau_n)}\\
        \int \mathcal{D}[\bar{\phi}\phi]e^{-s_k}=\prod_{n}\bigg(\sqrt{\det (-\partial_{\tau_n} +\epsilon_l+R_{k})}\bigg)^{-1}\\
        \int \mathcal{D}[\bar{\phi}\phi]e^{-s_k}\partial_{\ln k} s_k=\frac{1}{2}e^{-\frac{1}{2}\sum_{n}\ln \det (-\partial_{\tau_n} +\epsilon_l+R_{k})}\det\frac{\partial_{\ln k} R_k}{ (-\partial_\tau +\epsilon_l+R_{k})}\\
        \int \mathcal{D}[\bar{\phi}\phi]e^{-s_k}\partial_{\ln k} s_k=-\frac{1}{2}e^{-\frac{1}{2}\sum_{n}\ln \det (-\partial_\tau +\epsilon_l+R_{k})}\int \mathcal{D}[\bar{\phi}\phi] e^{\bar{\phi}\frac{\partial_{\ln k} R_k}{ (-\partial_\tau +\epsilon_l+R_{k})}\phi}\\
        \partial_{\ln k} s_k=-\frac{1}{2}Tr\bigg( \frac{\partial_{\ln k} R_k}{ (-\partial_\tau +\epsilon_l+R_{k})}\bigg)
     \end{split}
\end{equation}
Where we can readily see that $-\partial+\epsilon$ is nothing but the correlation function 
The boundary condition for bosons will be periodic for fermions it will be antiperiodic,

\begin{equation}
    \begin{split}
        \phi(\tau)=
        \begin{cases}
       \phi(\tau+\beta \hbar), for \hspace{2mm}  \text{bosons}\\ 
        -\phi(\tau+\beta \hbar), for \hspace{2mm} \text{fermions} \\
        \end{cases}
    \end{split}
\end{equation}
At one loop it is very trivial to show the equivalence between the two formalism. Now we go at the arbitrary point and integrate out the bosons at that point. If we introduce the vertex operator as the following to begin the 1-loop RG 
\begin{equation}
\begin{split}
    \hat{V}(\tau_n,\tau_{n-1})=\frac{1}{\det(R)}\int \prod_{i=n,n-1}\mathcal{D}[\bar{\phi}_{\tau_i}\phi_{\tau_{i}}] e^{-\sum_{i}\big(R_{i,i}\bar{\phi}_{\tau_i}\phi_{\tau_{i}}\big)-R_{n,n-1}\bar{\phi}_{\tau_n}\phi_{\tau_{n-1}}-R_{n-1,n}\bar{\phi}_{\tau_{n-1}}\phi_{\tau_{n}}} \\
    \det(R)=R_{n,n}R_{n-1,n-1}-R_{n,n-1}R_{n-1,n}
\end{split}
\end{equation}
\begin{equation}
    \begin{split}
        \mathcal{S}^{(1)}_{eff}=\frac{1}{\det(\mathcal{G}^{-1}_n+R)}\sum_{n} \bar{\phi}_{\tau_{n-1}}\bigg((\mathcal{G}^{-1}_{n-1}+R_{n-1,n-1})-\frac{R_{n,n-1}R_{n-1,n}}{(\mathcal{G}^{-1}_{n}+R_{n,n})}\bigg)\phi_{\tau_{n-1}}+\hat{V}(\tau_n,\tau_{n-1})
    \end{split}
\end{equation}
Using the translation property of the bosons we can write the above action as the following,
\begin{equation}
    \begin{split}
        \mathcal{S}^{(1)}_{eff}=\frac{1}{\det(\mathcal{G}^{-1}_n+R)}\sum_{n} \bar{\phi}_{\tau_{n}}\bigg((\mathcal{G}^{-1}_{n}+R_{n,n})-\frac{R_{n+1,n}R_{n,n+1}}{(\mathcal{G}^{-1}_{n+1}+R_{n+1,n+1})}\bigg)\phi_{\tau_{n}}+\hat{V}(\tau_{n+1},\tau_{n})
    \end{split}
\end{equation}
The second loop integrating out the (n+1) variables pairwise and using the transnational symmetry we get the following,
\begin{equation}
    \begin{split}
        \mathcal{S}^{(1)}_{eff}=\sum_{n} \bar{\phi}_{\tau_{n}}\bigg((\mathcal{G}^{-1}_{n}+R_{n,n})-\frac{R_{n+1,n}R_{n,n+1}}{(\mathcal{G}^{-1}_{n+1}+R_{n+1,n+1})}-\frac{R_{n,n+1}R_{n+1,n}}{(\mathcal{G}^{-1}_{n}+R_{n,n})-\frac{R_{n+1,n}R_{n,n+1}}{(\mathcal{G}^{-1}_{n+1}+R_{n+1,n+1})}}\bigg)\phi_{\tau_{n}}\\
        +\hat{V}(\tau_{n+2},\tau_{n+1})
    \end{split}
\end{equation}
Amputed correlation function from 1-loop can be written as the Jacobi's Continued fractions,
\begin{equation}
    \begin{split}
        \tilde{\mathcal{G}}^{-1}_n=(\mathcal{G}^{-1}_{n}+R_{n,n})-\frac{R_{n+1,n}R_{n,n+1}}{(\mathcal{G}^{-1}_{n+1}+R_{n+1,n+1})}-\frac{R_{n,n+1}R_{n+1,n}}{(\mathcal{G}^{-1}_{n}+R_{n,n})-\frac{R_{n+1,n}R_{n,n+1}}{(\mathcal{G}^{-1}_{n+1}+R_{n+1,n+1})}}
    \end{split}
\end{equation}
Now again introducing the vertex operator and carrying out the above derivation we get the following,
\begin{equation}
    \begin{split}
        \tilde{\mathcal{G}}^{-1(2)}_n=(\mathcal{G}^{-1}_{n}+R_{n,n})-\frac{R_{n+1,n}R_{n,n+1}}{(\tilde{\mathcal{G}}^{-1}_{(n\to n+1)+1}+R_{n+1,n+1})}\\
        -\frac{R_{n,n+1}R_{n+1,n}}{(\tilde{\mathcal{G}}^{-1}_{n\to n+1}+R_{n,n})-\frac{R_{n+1,n}R_{n,n+1}}{(\tilde{\mathcal{G}}^{-1}_{(n\to n+1)+1}+R_{n+1,n+1})}}
    \end{split}
\end{equation} 
This basically show from the above both the formalisms are equivalent and taking the $\log\bigg(\sqrt{\det( \tilde{\mathcal{G}}^{-1(2)}_n)}\bigg)$ the n-point function flow equation contains the $n+1$ and $n+2$ correlation function hence the scalar field FRG is equivalent to $\tau$-RG in Grassmann action.The vertex function $R_{i,j}$ can be chosen in such a way $\tau \to \infty$  vertex function will vanish and $\tau \to 0$ it will be unity.
\begin{equation}
    R_{i,j}=e^{-\frac{1}{2}(\tau_i^2+\tau_j^2)}
\end{equation}
This above derivation may be represented diagrammatically as in the fig \ref{fig:tau_RG}.
\section{$\tau$-Permutation of g-numbers}
Now we discrete the action but instead of introducing vertex operator we just expand the action in partition function by taking various permutation at the vertices. Again we can use  functional integration to find effective action and comparing the terms from operator structure derive the flow equations recurrence relations.
\begin{equation}
    \begin{split}
        \mathcal{S}=\int dt \sum_{kn}\big(\bar{\phi}_{kn}(i\frac{\partial}{\partial t}+\omega_{k})\phi_{kn}+i\gamma(\bar{\phi}_{kn}+\phi_{kn})\big)
    \end{split}
\end{equation}
Let's integrate out the kn pair of variables,
\begin{equation}
    \begin{split}
        \mathcal{Z}=\int \prod_{kn}\mathcal{D}[\bar{\phi}_{kn}\phi_{kn}]e^{-\mathcal{S}}
    \end{split}
\end{equation}
Carrying out integration of one pair of variables gives $\mathcal{Z}_{eff}=\int \prod_{kn-1}[\bar{\phi}_{kn-1}\phi_{kn-1}]e^{-\mathcal{S}_{eff}}$(using bosonic algebra $[\bar{\phi}_{k},\phi_{k'}]=\delta_{kk'}$ )  we get the following,
\begin{equation}
    \begin{split}
    \mathcal{Z}_{eff}&=\int \prod_{kn-1}[\bar{\phi}_{kn-1}\phi_{kn-1}]\bigg(\mathcal{G}^{-1}_{kn}+\sum_{kn}\big(\bar{\phi}_{kn-1}(\mathcal{G}^{-1}_{kn-1}\mathcal{G}^{-1}_{kn}-\gamma^2_{kn})\phi_{kn-1} \\ &+i(-\gamma^2_{kn}\gamma_{kn-1}+\mathcal{G}^{-1}_{kn}\gamma_{kn-1})(\bar{\phi}_{kn-1}+\phi_{kn-1})\bigg)\\
     \therefore   \mathcal{S}_{eff}&=\int dt \sum_{kn}\big(\bar{\phi}_{kn-1}(\mathcal{G}^{-1}_{kn-1}-\frac{\gamma^2_{kn}}{\mathcal{G}^{-1}_{kn}})\phi_{kn-1} \\ &+i(\frac{\gamma^2_{kn}\gamma_{kn-1}}{\mathcal{G}_{kn}}+\gamma_{kn-1})(\bar{\phi}_{kn-1}+\phi_{kn-1})\big)
    \end{split}
\end{equation}
\begin{figure}
    \centering
    \includegraphics[scale=1.2]{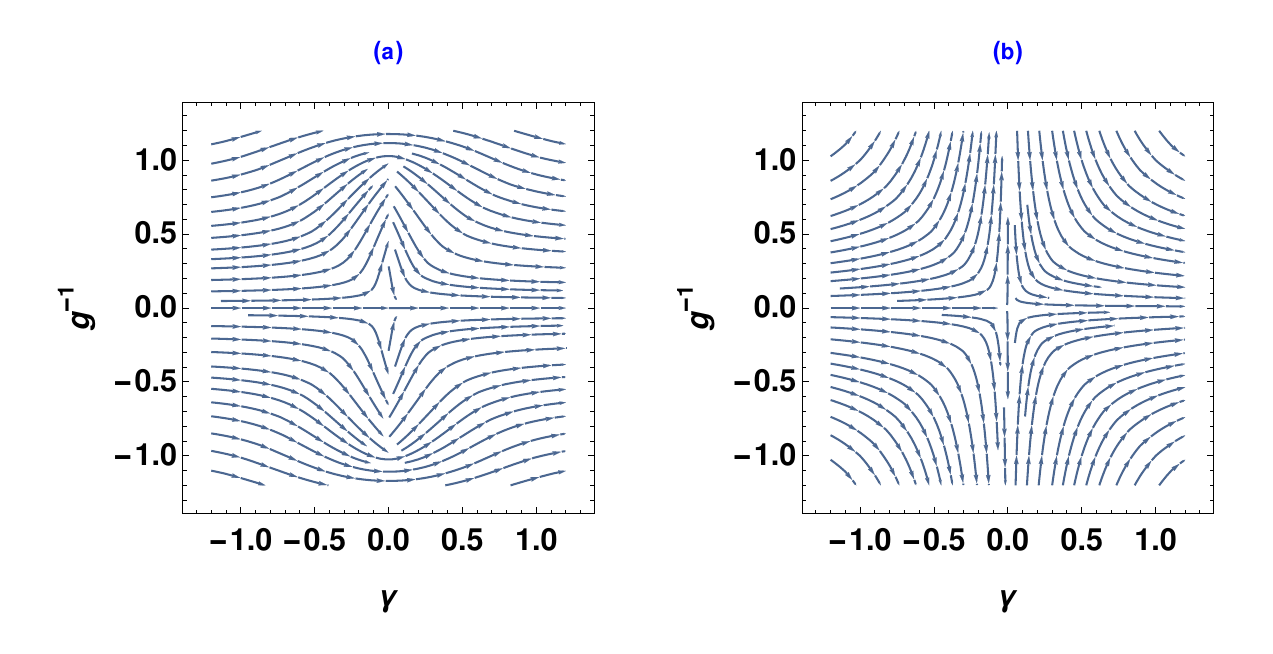}
     \caption{The panel labelled (a) is for smaller cutoff $\Lambda=1$  and panel (b) $\Lambda=100$. The cutoff plays a role in the transition as we can see from the RG flows there are unstable points when imaginary interaction $\gamma<0,Re(E_k)=0$ and there are stable points for $\gamma>0,Re(E_k)=0$ indicating the transition, also the  ground state can flow to purely imaginary.}
    \label{fig:E_gama}
\end{figure}

This will give us the following RG equations in recursive form at one step,
\begin{equation}
\label{eq:RG_equation}
    \begin{split}
        \mathcal{G}^{-1}_{kn}=\mathcal{G}^{-1}_{kn-1}-\frac{\gamma^2_{kn}}{\mathcal{G}^{-1}_{kn}}\\
         \gamma_{n}=\gamma_{n-1}+\frac{\gamma^2_{kn}\gamma_{kn-1}}{\mathcal{G}^{-1}_{kn}}
    \end{split}
\end{equation}
More general form of the above can be written as the continued fractions or as a product of the couplings at $n^{th}$- step as the following,
\begin{equation}
    \begin{split}
        \mathcal{G}^{-1}_{n}=\mathcal{G}^{-1}_{n-1}+\sum_{n',m'}\prod_{n',m'}\gamma_{n'}\mathcal{G}^{-1}_{m'}+\sum_{n'}\prod^{-\frac{\beta}{2}}_{n'<(n-2)} \mathcal{G}^{-1}_{n'}+\sum_{n'}\prod^{-\frac{\beta}{2}}_{n'<(n-2),even}(i)^{n'}\gamma_{n'}\\
        \gamma_{n}=\gamma_{n-1}+\sum_{n',m'}\prod_{(n',m')<(n-2),(odd,even)}(i)^{n'}\gamma_{n'}\mathcal{G}^{-1}_{m'}+\sum_{n'}\prod_{n'}\gamma_{n'}
    \end{split}
\end{equation}
\section{Lemma in normal ordered operators binomial expansion}
The various normal ordering identities are used in different contexts\cite{blasiak2007combinatorics,wyss2017noncommutative,kuchment2019some,fujii2004new}, One of  The key lemma states that if A and B determine a weyl algebra such that $[A,B]=1$ normal ordering of $(A+B)^n$ follows as,\\

\begin{equation}
    \begin{split}
        &(A+B)^{n}=\sum^{n}_{m=0}\sum^{min\lbrace m,n-m\rbrace}_{l=0}\begin{Bmatrix}n\\m\end{Bmatrix}_{l}B^{m-l}A^{n-m-l}\\
        &(\phi+\bar{\phi})^{n}=\sum^{n}_{m=0}\sum^{min\lbrace m,n-m\rbrace}_{l=0}\begin{Bmatrix}n\\m\end{Bmatrix}_{l}\bar{\phi}^{m-l}\phi^{n-m-l}\\
        &where \hspace{2mm}\begin{Bmatrix}n\\m\end{Bmatrix}_{l}=\frac{n!}{2^ll!(n-l)!(n-m-l)!}
    \end{split}
\end{equation}
This lemma can be proved by combinatorial way which is already attempted by many people earlier.We use this in renarmalization context to get the ground state correction.
 
\begin{figure}
    \centering
    \includegraphics[scale=1.3]{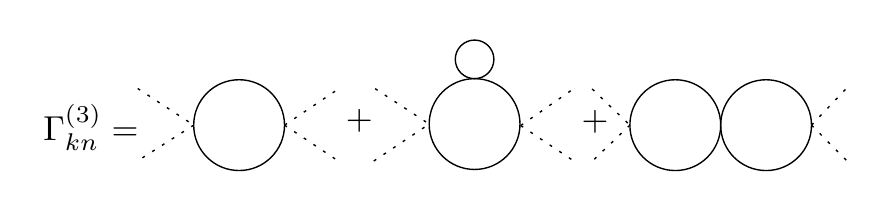}
     \caption{The third order contribution to the scale dependent action in conventional perturbation.}
    \label{fig:flow_E_gama}
\end{figure}
Expanding the above diagrams for 3 vertex and each of the will scaled by determinant, the determinant hitting zero is the special point we analyze it separately,
\begin{equation}
    \begin{split}
        diag^3_1=\frac{1}{det(\tau_n)}\bigg(\mathcal{G}^{-1}:\bar{\phi}\phi:\mathcal{G}^{-1}:\bar{\phi}\phi:i\gamma:\bar{\phi}+\phi:\bigg)+permutations \\
        diag^3_2=\frac{-1}{det(\tau_n)}\bigg(\mathcal{G}^{-1}:\bar{\phi}\phi:\gamma^2:\bar{\phi}+\phi::\bar{\phi}+\phi:\bigg)+permutations\\
        diag^3_2=\frac{1}{det(\tau_n)}\bigg(i\gamma:\bar{\phi}+\phi:i\gamma:\bar{\phi}+\phi:i\gamma:\bar{\phi}+\phi:\bigg)+permutations
    \end{split}
\end{equation}
We calculate these diagrams and get correction upto 3rd order as following,
\begin{equation}
    \begin{split}
        \delta\mathcal{G}^{-1}_{n}=-\frac{\gamma^2_{n}\mathcal{G}^{-1}-(\mathcal{G}^{-1})^2_n-(\mathcal{G}^{-1})^3_{n}}{(\mathcal{G}^{-1})_{n}-\gamma^2_{n}}\\
        \delta\gamma_{n}=-\frac{-\gamma_{n}\mathcal{G}^{-1}_{n}+\gamma^3_n-\gamma_n(\mathcal{G}^{-1})^2_n}{\mathcal{G}^{-1}_{n}-\gamma^2_{n}}\\
        \implies \frac{\delta \mathcal{G}^{-1}_{n}}{\delta \gamma_{n}}=\frac{\mathcal{G}^{-1}_{n}}{\gamma_n},\hspace{2mm} \log\frac{\mathcal{G}^{-1}}{\gamma}=invarient
    \end{split}
\end{equation}
The RG flow for the integer power is still the same in terms of seperatrix , but equal time permutations does not renormalize the exceptional points. The above  shows if we don't consider higher fluctuations to nth $\tau$ vertex then the perturbative RG will not capture the correct physics.As it shows the from third order calculations we need higher loop corrections.
\section{RG in $\tau$-space}
There are various methods in literature performing renormalization effectively in all the energy scales. Our interest in this work is to find the characteristic temperature scale for BEC to occur and explore the phase beyond the realm of real eigenvalues. Since we can use the connection between the Matsubara points by defining fourier transform as $\phi(\tau)=\frac{1}{\beta}\sum_{n}e^{-i\omega_n\tau}\phi(\omega_n)$ where  $\omega_n=\frac{2n\pi}{\beta}$ 
\begin{figure}
    \centering
    \includegraphics[scale=1.2]{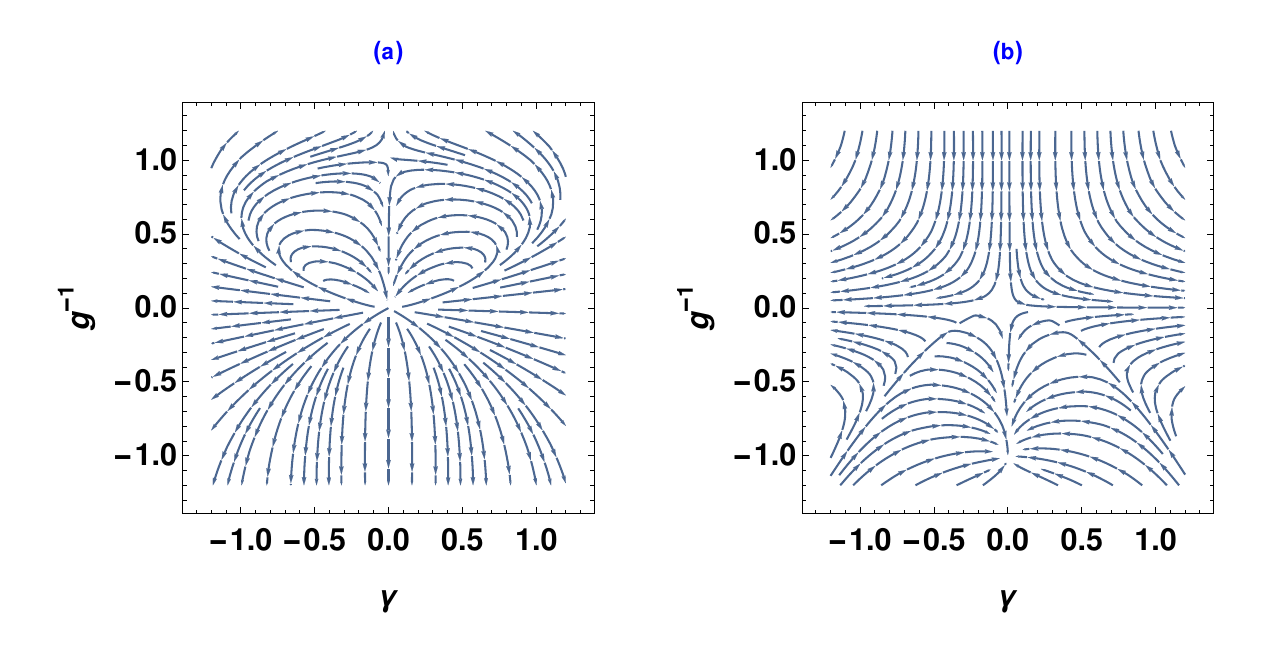}
     \caption{The panel labelled (a) is for smaller cutoff $\Lambda=1$  and panel (b) $\Lambda=100$. The cutoff plays a role in the transition as we can see from the RG flows there are unstable points when imaginary interaction $\gamma<0,Re(E_k)=0$ and there are stable points for $\gamma>0,Re(E_k)=0$ indicating the transition, also the  ground state can flow to purely imaginary.}
    \label{fig:flow_E_gama}
\end{figure}
We can schematically represent the RG method in the Matsubara points as the following,

\subsection{Analytic Solution of RG equations to 1 loop}
The recursion equation is eq \ref{eq:RG_equation} can be used to set the ODE to solve analytically as the following,
\begin{equation}
    \begin{split}
        (-\frac{\partial}{\partial\tau_n}+\epsilon)\tilde{\mathcal{G}}_{n}+\tilde{\mathcal{G}}_n\mathcal{G}_{n-1}= \mathcal{G}^2_{n-1} +(\frac{\partial}{-\partial\tau_n}+\epsilon)\mathcal{G}_{n-1}-\gamma^2_n\\
        (-\frac{\partial}{\partial\tau_n}+\epsilon)\tilde{\gamma}_{n}+\tilde{\gamma}_n\mathcal{G}_{n-1}= \gamma_{n-1}\mathcal{G}_{n-1} +(-\frac{\partial}{\partial\tau_n}+\epsilon)\gamma_{n-1}-\gamma^2_n\gamma_{n-1}\\
         -\frac{\partial}{\partial\tau_n}\tilde{\mathcal{G}}_{n}= \mathcal{G}^2_{n-1} -\gamma^2_n+(n,n-1)terms \\
         (-\frac{\partial}{\partial\tau_n})\tilde{\gamma}_{n}= \gamma_{n-1}\mathcal{G}_{n-1} -\gamma^2_n\gamma_{n-1}+(n,n-1)terms
    \end{split}
\end{equation}
Using the initial condition $\gamma_n=\gamma_{n-1}=\gamma$ and $\mathcal{G}^{-1}_{n}=\mathcal{G}^{-1}_{n-1}=\mathcal{G}^{-1}$we get following equations, and the $(n,n-1)$ terms will be taken care at the 2-loop RG 
\begin{equation}
    \begin{split}
        -\frac{\partial}{\partial\tau_n}\tilde{\mathcal{G}}_{n}= \mathcal{G}^2 -\gamma^2 \\
         -\frac{\partial}{\partial\tau_n}\tilde{\gamma}_{n}= \gamma\mathcal{G} -\gamma^3
    \end{split}
\end{equation}
The above can be solved analytically  with substitution $\frac{\gamma^3}{(\mathcal{G}^{-1})^2}=t$
\begin{equation}
    \begin{split}
        \sqrt{\frac{\gamma}{t}}-\frac{1}{3}\frac{\gamma^{\frac{3}{2}}}{t^{\frac{3}{2}}}\frac{dt}{d\gamma}=\frac{1-\gamma t}{\sqrt{\frac{t}{\gamma}}-t},\hspace{2mm}\therefore\hspace{2mm}
        1-\frac{1}{3}\frac{\gamma}{t}\frac{dt}{d\gamma}=\frac{1-\gamma t}{1-\sqrt{t\gamma}}\\
        \implies -\frac{1}{3}\frac{\gamma}{t}\frac{dt}{d\gamma}=\frac{\sqrt{t\gamma}-\gamma t}{1-\sqrt{t\gamma}}=\sqrt{t\gamma},\hspace{2mm} -\int\frac{1}{3}\frac{1}{t^{\frac{3}{2}}}dt=\int\frac{1}{\sqrt{\gamma}}d\gamma+C\\
        \frac{2}{3}\frac{1}{\sqrt{t}}=2\sqrt{\gamma}+C,\hspace{2mm} \implies \frac{2}{3}\mathcal{G}^{-1}=2\gamma^{2}+C\gamma^{\frac{3}{2}}
    \end{split}
\end{equation}

Integral constant carries useful information for example if the both coupling scale as unity then $C= 2/3- 2\sqrt{\gamma}$ and for spectral function reversing sign we have $C=-i\frac{2}{3}\frac{1}{\gamma^{3/4}}-2\sqrt{\gamma}$ which are the two regimes as $\mathcal{G}^{-1}=\gamma^{3/2}$ and  $\mathcal{G}^{-1}=-i\gamma^{3/2}$.We get the scale for the problem in these two phases as $\beta=-\frac{3a \, _2F_1\left(-\frac{1}{2},1;\frac{1}{2};-a(\mathcal{G}^{-1})^{2/3}\right)}{\sqrt[3]{\mathcal{G}^{-1}}}$ and $\beta=-\frac{2b}{3 \gamma^{3/2}}-\frac{1}{\gamma}-\frac{2b}{\sqrt{\gamma}}+\log (\gamma)-2 \log \left(1-\sqrt{\gamma}\right)$.The constants $(a,b)$ in the scale are unity in conventional boson phase and for symmetry broken phase they tend to $((-1)^{1/3},i)$ which imply the complex scale in the other phase. 
\begin{figure}
    \centering
    \includegraphics[scale=0.7]{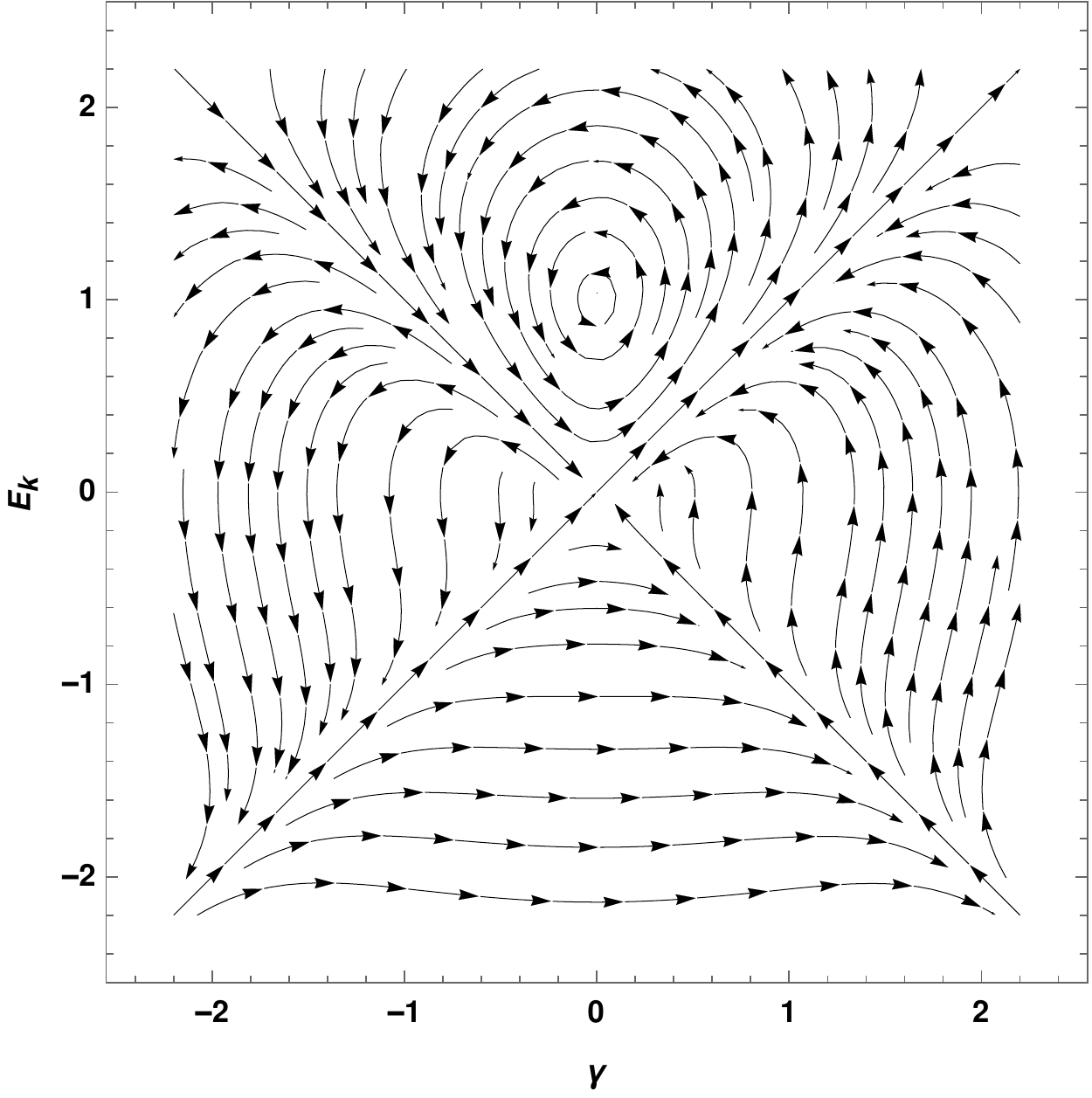}
     \caption{The flow of effective ground state with complex potential parameter for the generator as worked out in the appendix we observe there is limit cycle in the RG flow}
    \label{fig:flow_SO1}
\end{figure}
\subsection{Generalization to N-power  potential}
If we carry out the procedure for n-tau integration for general N-power potental we get the following RG equations at one loop,
\begin{equation}
    \begin{split}
        \frac{d\mathcal{G}^{-1}}{d\tau}=(\mathcal{G}^{-1})^2-N^2\gamma^{2N}\\
       N\gamma^{N-1} \frac{d\gamma}{d\tau}=N\gamma^{N}\mathcal{G}^{-1}
    \end{split}
\end{equation}
\begin{figure}
    \centering
    \includegraphics[scale=1.0]{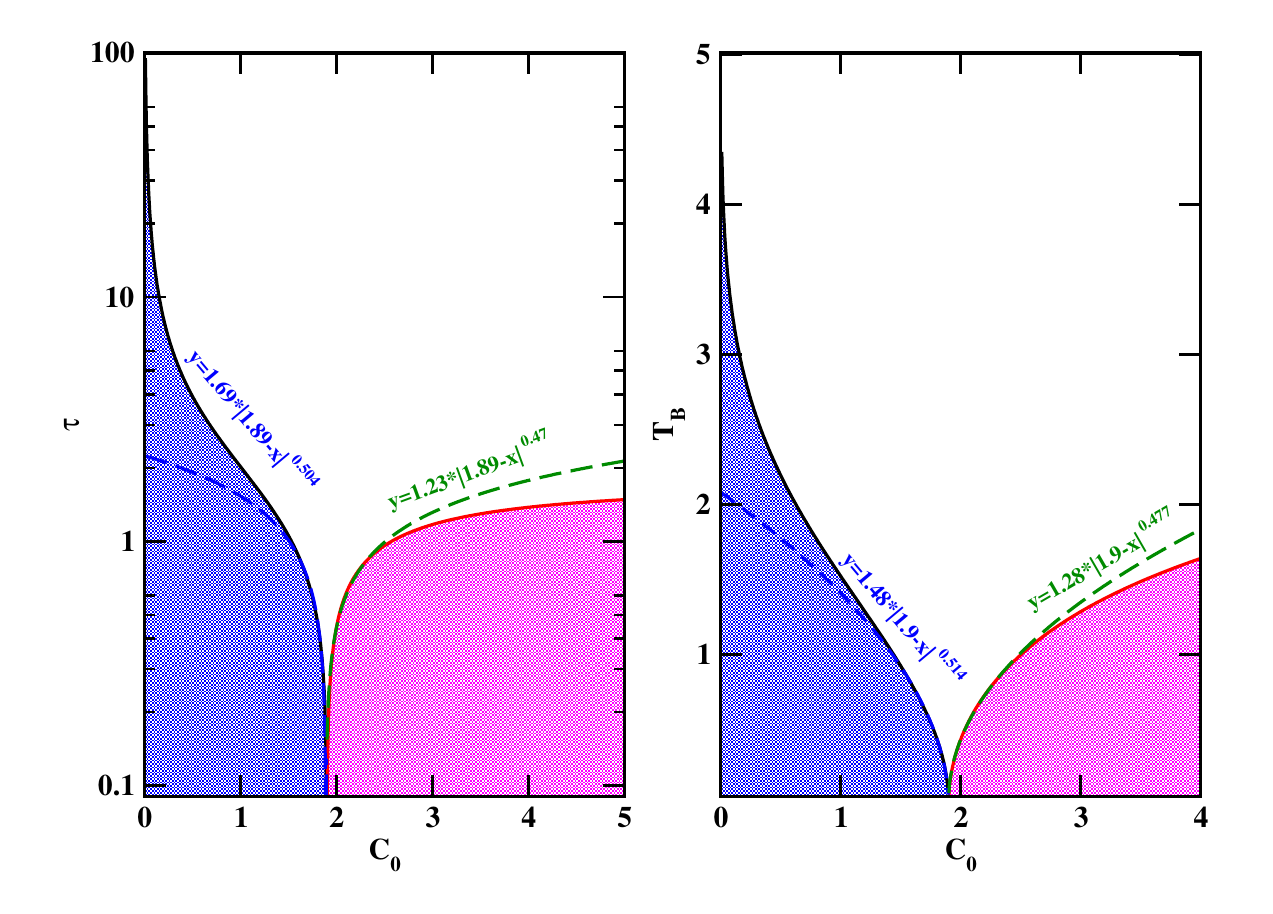}
     \caption{Left pannel is the renormallization of flow parameter Matsubara time(y-axis) which can be analytic continuation, Right pannel is the temperature scale for bosons after the Matsubara sum(y-axis) and both pannel x-axis is a scale $C_0=\log{\big(e^{\frac{g^2}{\gamma^2}}\frac{1}{\gamma}\big)}$}
    \label{fig:flow_SO2}
\end{figure}
\section{RG in left-right Basis}
Let's re write the action in the new basis as the following,
\begin{equation}
    \begin{split}
        \mathcal{S}=&\int d\tau\bigg(\bar{\phi}_{R}\mathcal{G}^{-1}\phi_{R}-\bar{\phi}_{R}\mathcal{G}^{-1}\phi_{L}-\bar{\phi}_{L}\mathcal{G}^{-1}\phi_{R} +\bar{\phi}_{R}\mathcal{G}^{-1}\phi_{R}+i^N\gamma^N(\bar{\phi}_R-\bar{\phi}_L+\phi_{R}-\phi_{L})^N\bigg)
    \end{split}
\end{equation}
We can rewrite the action in the L-R basis by scaling the g-numbers and their corresponding commutation algebra as the following(please note that we did not scale the constant phase and independent of k as well for simplicity),
\begin{equation}
    \begin{split}
        \phi_L=e^{-i\frac{\nu}{N}}\phi,\phi_R=e^{i\frac{\nu}{N}}\phi
        \implies [\bar{\phi}_{Lk},\phi_{Rk'}]=e^{\frac{2i\nu}{N}}\delta_{kk'}\\ [\bar{\phi}_{Rk},\phi_{Lk'}]=e^{-\frac{2i\nu}{N}}\delta_{kk'},
        [\bar{\phi}_{\eta k},\phi_{\eta k'}]=\delta_{kk'}
    \end{split}
\end{equation}
Let $\nu$ is the arbitrary phase which will scale the field variables and action in this basis will be,
\begin{equation}
    \begin{split}
        \tilde{\mathcal{S}}=&\int d\tau\bigg(\bar{\phi}\big(\mathcal{G}^{-1}(1-\cos{\frac{2\nu}{N}})\big)\phi+e^{\frac{iN\pi}{2}}\gamma^N\bigg[\sin{\frac{\nu}{N}}\bigg]^N(\bar{\phi}-\phi)^N\bigg)\\
        =&\int d\tau\bigg(\bar{\phi}\big(2\mathcal{G}^{-1}\sin^{2}{\frac{\nu}{N}}\big)\phi+e^{\frac{iN\pi}{2}}\gamma^N\bigg[\sin{\frac{\nu}{N}}\bigg]^N(\bar{\phi}-\phi)^N\bigg)
    \end{split}
\end{equation}
We can write the path integral in the Matsubara time 
\begin{equation}
    \begin{split}
        \mathcal{Z}=&\int\prod_{n}[d\bar{\phi}_{n}d\phi_n]e^{-\bigg(\bar{\phi}_n\big(2\mathcal{G}^{-1}\sin^{2}{\frac{\nu}{N}}\big)\phi_n\bigg)}\prod^{N}_{m=0}\prod^{min(m,N-m)}_{l=0}e^{-\bigg(\begin{Bmatrix}N\\m\end{Bmatrix}_{l}e^{\frac{iN\pi}{2}}(\gamma)^N\sin^N{\frac{\nu}{N}}\bar{\phi}^{m-l}_n\phi^{N-m-l}_n\bigg)}
    \end{split}
\end{equation}
We show explicit calculation on getting one-loop RG equations in appendix,Also we can see from operator structure the contribution from $x^N$ potential can only appear for $m-l=N-m-l$ which basically means $m=N/2$ term in the time orderd expansion is going to contribute to the diagonal Matsubara-g.
\begin{equation}
    \begin{split}
        \frac{d\tilde{\mathcal{G}}^{-1}}{d\tau}=(\tilde{\mathcal{G}}^{-1})^2-N^2\gamma^{2N}\sin^{2N}{\frac{\nu}{N}}\\
       N\gamma^{N-1}\frac{d\gamma}{d\tau}=(-\gamma)^N\sin^{2}{\frac{\nu}{N}} \mathcal{G}^{-1}
    \end{split}
\end{equation}
We can do separation of variables as the following,
\begin{equation}
    \begin{split}
         (\tilde{\mathcal{G}}^{-1})^{2N}\frac{d\tilde{\mathcal{G}}^{-1}}{d\tau}-(\tilde{\mathcal{G}}^{-1})^{2N+2} 
         =(-1)N^{(2N+2)}\big(\frac{d\gamma}{d\tau}\big)^{\frac{1}{2N}}\sin^{2N}{\frac{\nu}{N}}=k\\
         \implies \int \frac{(\tilde{\mathcal{G}}^{-1})^{2N}}{k+(\tilde{\mathcal{G}}^{-1})^{2N+2}} d(\tilde{\mathcal{G}}^{-1})=\int^{\frac{\beta}{2}}_{-\frac{\beta}{2}} d\tau
    \end{split}
\end{equation}
Solution after integrating the  above RG equations are as follows,
\begin{equation}
    \begin{split}
    \label{eq:RGsol}
      \beta=-\frac{(\tilde{\mathcal{G}}^{-1})^{2 N+1} \, _2F_1\left(1,\frac{2 N+1}{2 N+2};\frac{4 N+3}{2 N+2};-\frac{(\tilde{\mathcal{G}}^{-1})^{2 N+2}}{k}\right)}{2 k N+k}\\
      \tilde{\gamma}=\frac{\beta k^{\frac{1}{2N}}N^{-\frac{(2N+2)}{(2N)}}}{\sqrt{\sin{\frac{\nu}{N}}}}\propto N^{-1/N-1/2}\hspace{2mm} for \hspace{2mm} \nu \hspace{2mm} small \\
     where\hspace{2mm} \tilde{\mathcal{G}}^{-1}=\sum_{n}(\partial_{\tau_{n}}+E_{k})\sin^{2}{\frac{\nu}{N}}
    \end{split}
\end{equation}
If we convert the retarded $\mathcal{G}^{-1}$ RG equation to $g$ then we get the exponent as the coordinate of the confluent geometric function,
\begin{equation}
    \begin{split}
       \beta= \mathcal{G} \, _2F_1\left(1,\frac{1}{2 N};1+\frac{1}{2 N};k \mathcal{G}^{2 N}\right)
    \end{split}
\end{equation}
From the above equation (\ref{eq:RGsol}) we recover the exponent for 1D BEC(when the phase is very small) derived earlier in various context \cite{bayindir1999bose,bagnato1991bose}.  
\section{General n-point Tau FRG}
Earlier sections we have seen how the imaginary time RG is analytically tractable for non-trivial phase of the wave-function and performed at finite temperatures the expansion of the RG equations take interesting mathematical forms which is one of the major result of this work.
\subsection{Saddle Point Solution of Action}
Effective action derived after integrating out pair of variables at the arbitrary vertex we can use the condition of invariance as taking the variation of action with the phase factor introduced and look for where the saddle point solutions can exist, ie $\frac{\delta\mathcal{S}}{\delta \nu}=f'(\nu)\mathcal{S}_{eff}+f(\nu)\mathcal{S}'_{eff}=0$  
\begin{equation}
    \begin{split}
        \ln(\mathcal{S}_{eff})=\sin \left(\frac{v}{N}\right) \left((N-2) \, _2F_1\left(1,\frac{1}{2-N};1+\frac{1}{2-N};-\frac{\mathcal{G}^{-1} N \sin ^{2-N}\left(\frac{v}{N}\right)}{e^{-i\frac{N\pi}{2}} \gamma^N}\right)+2\right)\ln(\mathcal{S}_0)
    \end{split}
\end{equation}
The several minima of the theory can be found at the following points,
\begin{equation}
    \begin{split}
        \frac{\nu}{N}=n\pi +(-1)^n\arcsin\bigg(-\frac{2\mathcal{G}^{-1}e^{-i\frac{N\pi}{2}}}{N\gamma^N}\bigg)^{\frac{1}{N-2}} \hspace{2mm} \forall n\in\mathbb{N}
    \end{split}
\end{equation}
\section{One and Two body terms in grassmann}
Let's rewrite the Schrodinger equation in terms of the g-variables.

\begin{equation}
\label{eq:schroo}
    \begin{split}
    &\bigg(\sum_{\alpha \beta}T_{\alpha \beta}\phi^{*}_{\alpha}\frac{\partial}{\partial \phi^{*}_{\beta}}+\sum_{\alpha\beta\gamma\delta}\langle\alpha\beta|V|\gamma\delta\rangle\phi^{*}_{\alpha}\phi^{*}_{\beta}\frac{\partial}{\partial \phi^{*}_{\gamma}}\frac{\partial}{\partial \phi^{*}_{\delta}}\bigg)\psi(\phi^{*})=E\psi(\phi^{*})
    \end{split}
\end{equation}
If we do an exercise by analytic continuation of the Grassmann(g) numbers as $\tilde{\phi}_{\mathcal{K}}=\phi_{\mathcal{K}}+i\eta_{\mathcal{K}}$ where $\mathcal{K}=\alpha,\beta,\gamma,\delta$ and  we get the following,
\begin{equation}
    \begin{split}
         &\bigg(\sum_{\alpha \beta}\tilde{T}_{\alpha \beta}\tilde{\phi}^{*}_{\alpha}\frac{\partial}{\partial \tilde{\phi}^{*}_{\beta}}+\sum_{\alpha\beta\gamma\delta}\langle\alpha\beta|\tilde{V}|\gamma\delta\rangle\tilde{\phi}^{*}_{\alpha}\tilde{\phi}^{*}_{\beta}\frac{\partial}{\partial \tilde{\phi}^{*}_{\gamma}}\frac{\partial}{\partial \tilde{\phi}^{*}_{\delta}}\bigg)\psi(\phi^{*}+i\eta \big)=E\psi(\phi^{*}+i\eta)\\
         &=\bigg(\sum_{\alpha \beta}\tilde{T}_{\alpha \beta}\phi^{*}_{\alpha}\frac{\partial}{\partial \phi^{*}_{\beta}}-i\tilde{T}_{\alpha \beta}(\phi^{*}_{\alpha}\frac{\partial}{\partial \eta^{*}_{\beta}}+\eta^{*}_{\alpha}\frac{\partial}{\partial \phi^{*}_{\beta}})+\tilde{T}_{\alpha \beta}\eta^{*}_{\alpha}\frac{\partial}{\partial \eta^{*}_{\beta}} \\ &+\sum_{\alpha\beta\gamma\delta}\langle\alpha\beta|\tilde{V}|\gamma\delta\rangle\tilde{\phi}^{*}_{\alpha}\tilde{\phi}^{*}_{\beta}\frac{\partial}{\partial \tilde{\phi}^{*}_{\gamma}}\frac{\partial}{\partial \tilde{\phi}^{*}_{\delta}}\bigg)\psi(\phi^{*}+i\eta \big)
    \end{split}
\end{equation}
We expand two body term in appendix to show this will yield the complex interaction but at some special points it will be in the eigenstate probability conserved regime that we will analyze from coherent states.
Also from One body contribution we can see analytic continuation would yield same Hamiltonian back if the complex contribution vanish as the following,
\begin{equation}
    \begin{split}
        \sum_{\alpha \beta}\bigg(\phi^*_{\alpha}\frac{\partial}{\partial \eta^*_\beta}+\eta^*_\alpha\frac{\partial}{\partial \phi^*_\beta} \bigg)\psi=0
    \end{split}
\end{equation}
Let's do a variable seperable method and write $\psi=\xi(\phi^*)\chi(\eta^*)$ to find what is the eigenstate which satisfy this,
\begin{equation}
    \begin{split}
        \sum_{\alpha \beta}\frac{1}{\chi\eta^*_\alpha}\frac{\partial \chi}{\partial \eta^*_\beta}=-\sum_{\alpha \beta}\frac{1}{\xi\phi^*_{\alpha}}\frac{\partial \xi}{\partial \phi^*_\beta} =k\\
        \implies \xi=e^{k\int \phi^*_\alpha  d\phi^*_\beta}\xi_0,\hspace{2mm}\chi=\chi_0e^{-k\int \eta^*_\alpha d\eta^*_\beta}\\
    \therefore    \psi=e^{k\int \phi^*_\alpha  d\phi^*_\beta}\xi_0 \otimes \chi_0e^{-k\int \eta^*_\alpha d\eta^*_\beta}
    \end{split}
\end{equation}
As long as we can write the analytically continued states as pure product state we get the real eigenvalues and we get the unitary dynamics for the complex interaction.This state can be used to do perturbation expansion for the two body term to get the solution for many body systems.
Rewriting the coherent states in the analytically continued g-variables,
\begin{equation}
    \begin{split}
        \psi(\phi^{*}-i\eta^*)=\int\prod_{\alpha}\frac{d\phi'^*_\alpha d\phi'_\alpha}{2i\pi}e^{(\phi'^*_\alpha-\phi'_\alpha)(\phi^*_\alpha-i\eta^*)}\psi(\phi'^*_\alpha-i\eta'^*_\alpha)\\
        \int\prod_{\alpha}\frac{d\phi'^*_\alpha d\eta'_\alpha}{2i\pi}e^{(\phi'^*_\alpha-\phi'_\alpha)(\phi^*_\alpha-i\eta^*)}\psi(\phi'^*_\alpha-i\eta'^*_\alpha)\\
        \int\prod_{\alpha}\frac{d\eta'^*_\alpha d\phi'_\alpha}{2i\pi}e^{(\phi'^*_\alpha-\phi'_\alpha)(\phi^*_\alpha-i\eta^*)}\psi(\phi'^*_\alpha-i\eta'^*_\alpha)\\
        \int\prod_{\alpha}\frac{d\eta'^*_\alpha d\eta'_\alpha}{2i\pi}e^{(\phi'^*_\alpha-\phi'_\alpha)(\phi^*_\alpha-i\eta^*)}\psi(\phi'^*_\alpha-i\eta'^*_\alpha)
    \end{split}
\end{equation}
We will write the above functional integral in polar form 
\begin{equation}
    \begin{split}
        \int \frac{d\phi'^*_\alpha d\phi'_\alpha}{2i\pi}\to\int dydx\to\int dr d\theta
    \end{split}
\end{equation}
After analytically continuing the polar form can be written in the following way,
\begin{equation}
    \begin{split}
       & \int \frac{d\phi'^*_\alpha d\phi'_\alpha}{2i\pi}=\int (r+\eta_r)drd\theta+\int (r+\eta_r)d\eta_rd\theta +\int (r+\eta_r)drd\theta+\int(r+\eta_r) d\eta_rd\eta_\theta\\
     &\psi(\phi^*) = \int (r+\eta_r)dr'd\theta' e^{-2ir'\sin{\theta}\phi^*}\sum_{n}c_n r'^ne^{n\theta'}\\
      &=\sum_ne^{in\pi/2}\frac{\Gamma(n)c_n}{2^{n+1}}\oint\frac{e^{in\theta'}}{(\sin{\theta'})^n}d\theta'\frac{1}{(\phi^*-i\eta^*)^n}\\
      &=\sum_ne^{in\pi/2}\frac{\Gamma(n)c_n}{2^{n+1}}\bigg(\frac{1}{(n-1)!}\frac{d^{n-1}}{dz^{n-1}}\frac{z^{2n}}{(z+1)^n}\bigg)\frac{1}{(\phi^*-i\eta^*)^n}
    \end{split}
\end{equation}
Above consist of nth order pole 
This will yield the following for single particle,
\begin{equation}
\begin{split}
\psi(\phi^*-i\eta^*)=\sum_{n}\frac{\Gamma(n)C_{n}}{\phi^{*n}}
\end{split}
\end{equation}
These $C_n$ consist of the residues  as mentioned earlier.Also this satisfies the proposition-1 mentioned in earlier section.
\begin{figure}
    \centering
    \includegraphics[scale=0.7]{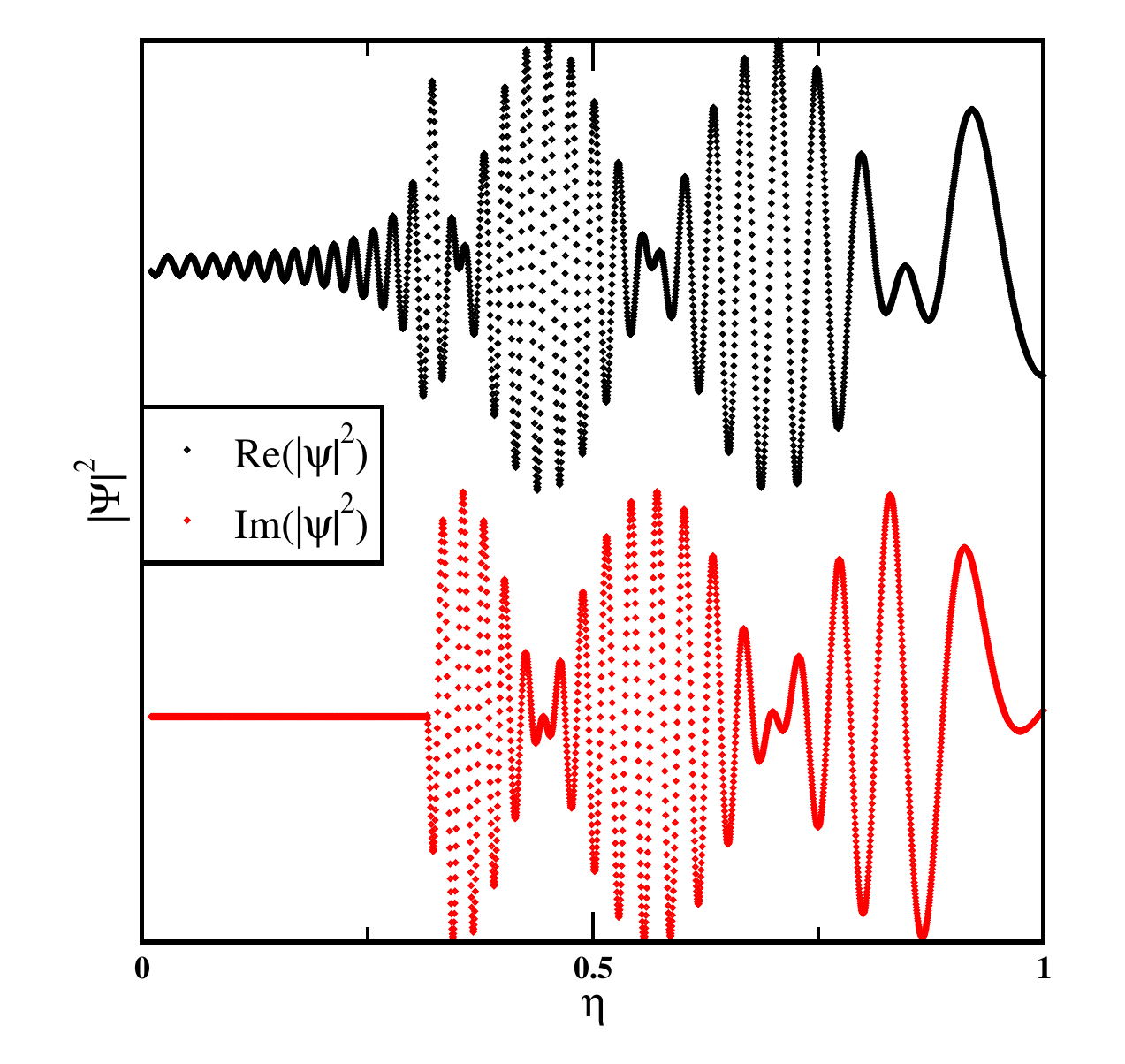}
     \caption{Analytically continued in complex plane for the bosonic Grassmann numbers and we can observe the imaginary part of the eigenstate probability(Red) and there will be real part of prabability  }
    \label{fig:flow_SO3}
\end{figure}
\begin{figure}
    \centering
    \includegraphics[scale=.5]{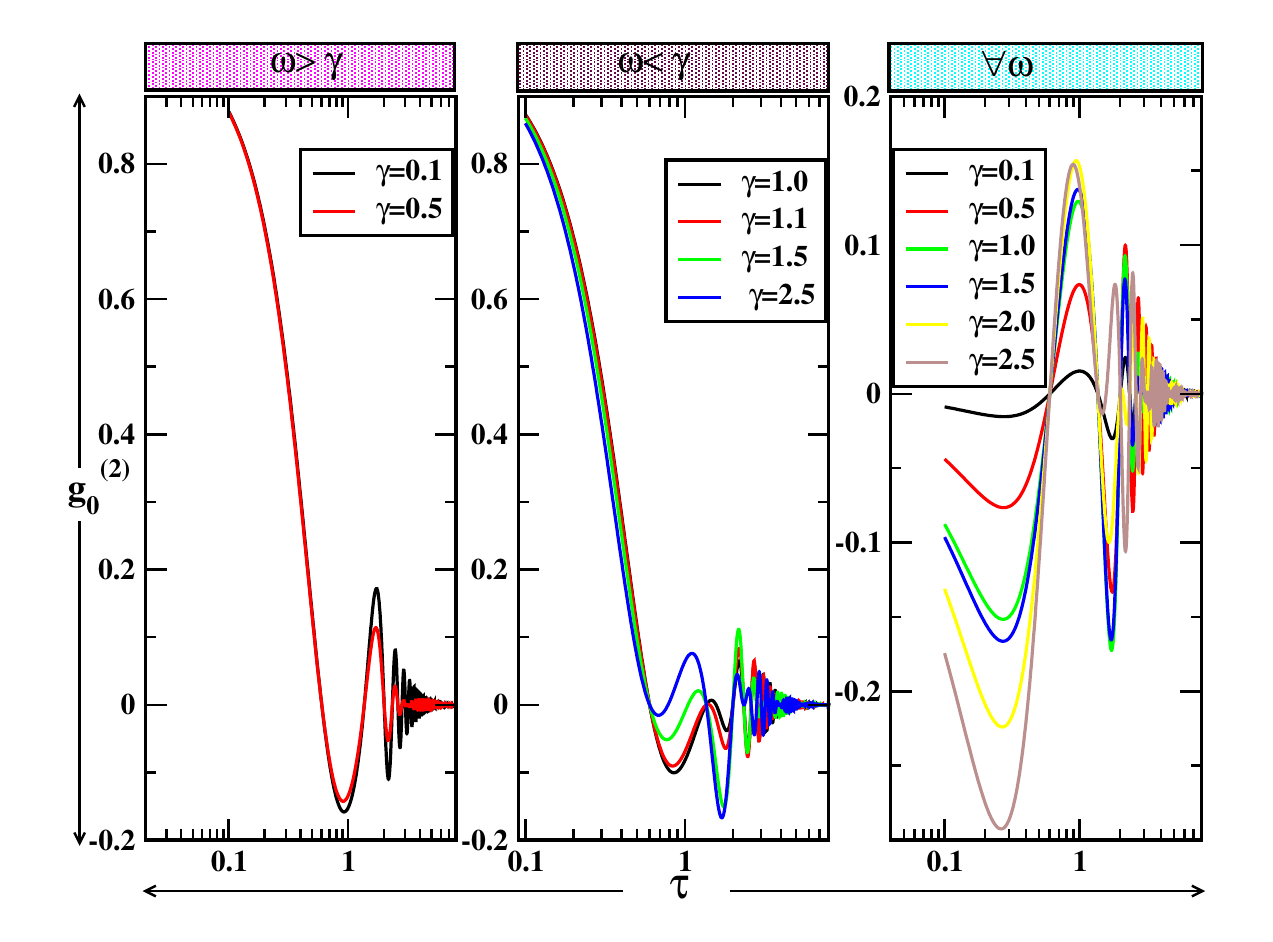}
     \caption{We compute the average boson number from 1-loop results ie., $\langle g^{(2)}_0 \rangle$ left panel labelled $\omega>\gamma$,$\omega<\gamma$ correspond to real part of the correlation function and right most (labelled $\forall \omega$) correspond to imaginary part.We notice significantly small oscillation of the imaginary weight for lesser $\gamma$ strength.}
    \label{fig:flow_SO4}
\end{figure}
\begin{figure}
    \centering
    \includegraphics[scale=.6]{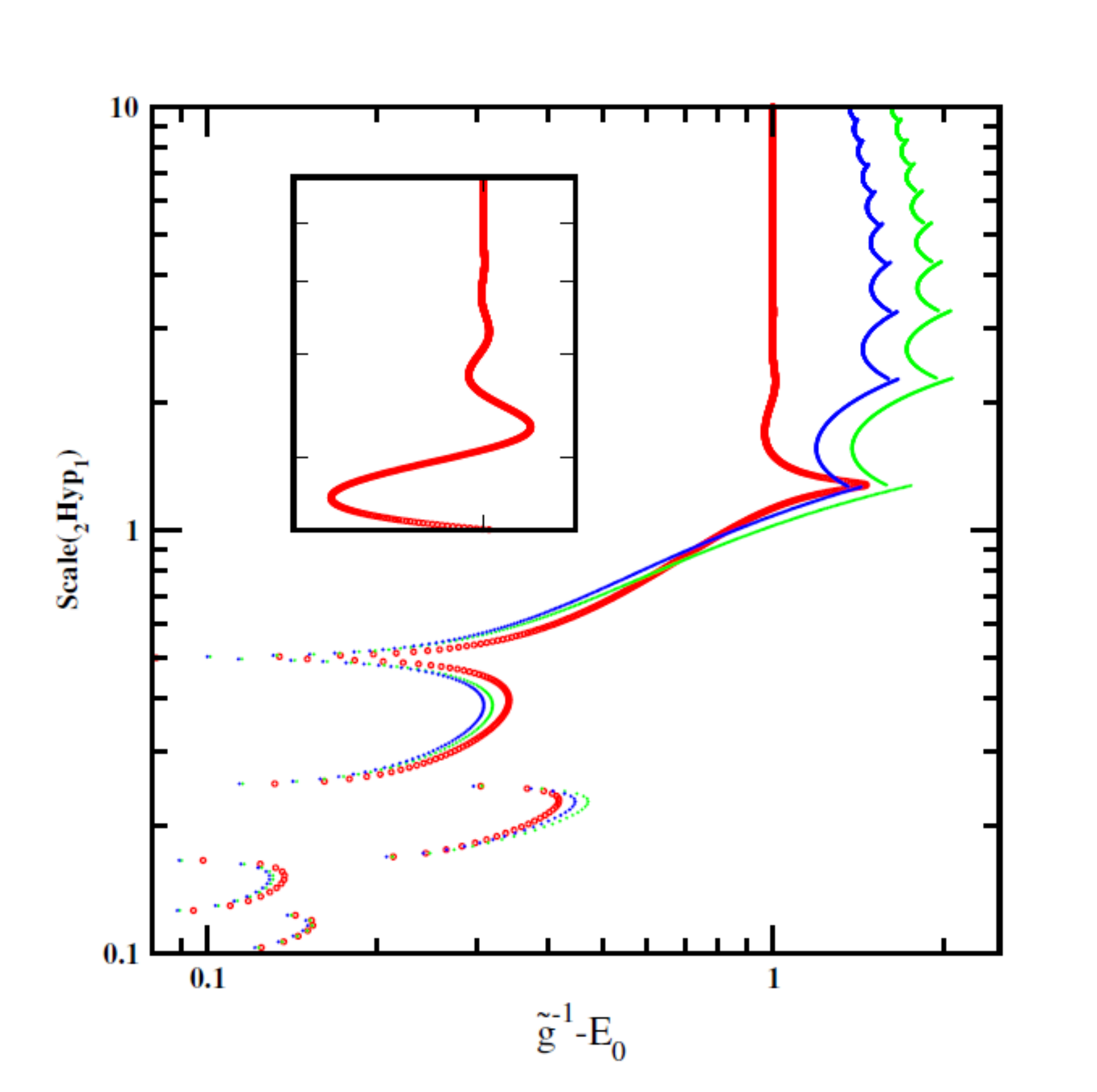}
     \caption{The RG Invarient (gauss-hypergeometric function) plotted with the ground state energy subtracted renormalized spectral function which is intricatly connected to the complex potential power nonetheless it shows the transition  and qualitative behaviours as that of the  results\cite{bender1999pt,bender2018asymptotic,bender2021pt,bender2004extension}.The red set of points are for $E_0 > \gamma$, blue set of ponts $E_0 = \gamma$ and $E_0<\gamma$ correspond to green data points.}
    \label{fig:flow_SO6}
\end{figure}
\section{Energy with N}
Earlier section we discussed how the RG is analytically tractable with closed form solutions and existence of limit cycles.The implication of wavefunction RG on the couplings can be readily seen through functional forms.The bare couplings of the theory brought back with some initial conditions the L-R RG equations are computed. We performed some elementary numerical study to see how well these RG results compare to the exact results from various methods like Wentzel–Kramers–Brillouin(WKB) and other perturbative methods as well. This also give some qualitative similar features of the groundstate energy with the complex potential power.\\
\begin{figure}
    \centering
    \includegraphics[scale=.5]{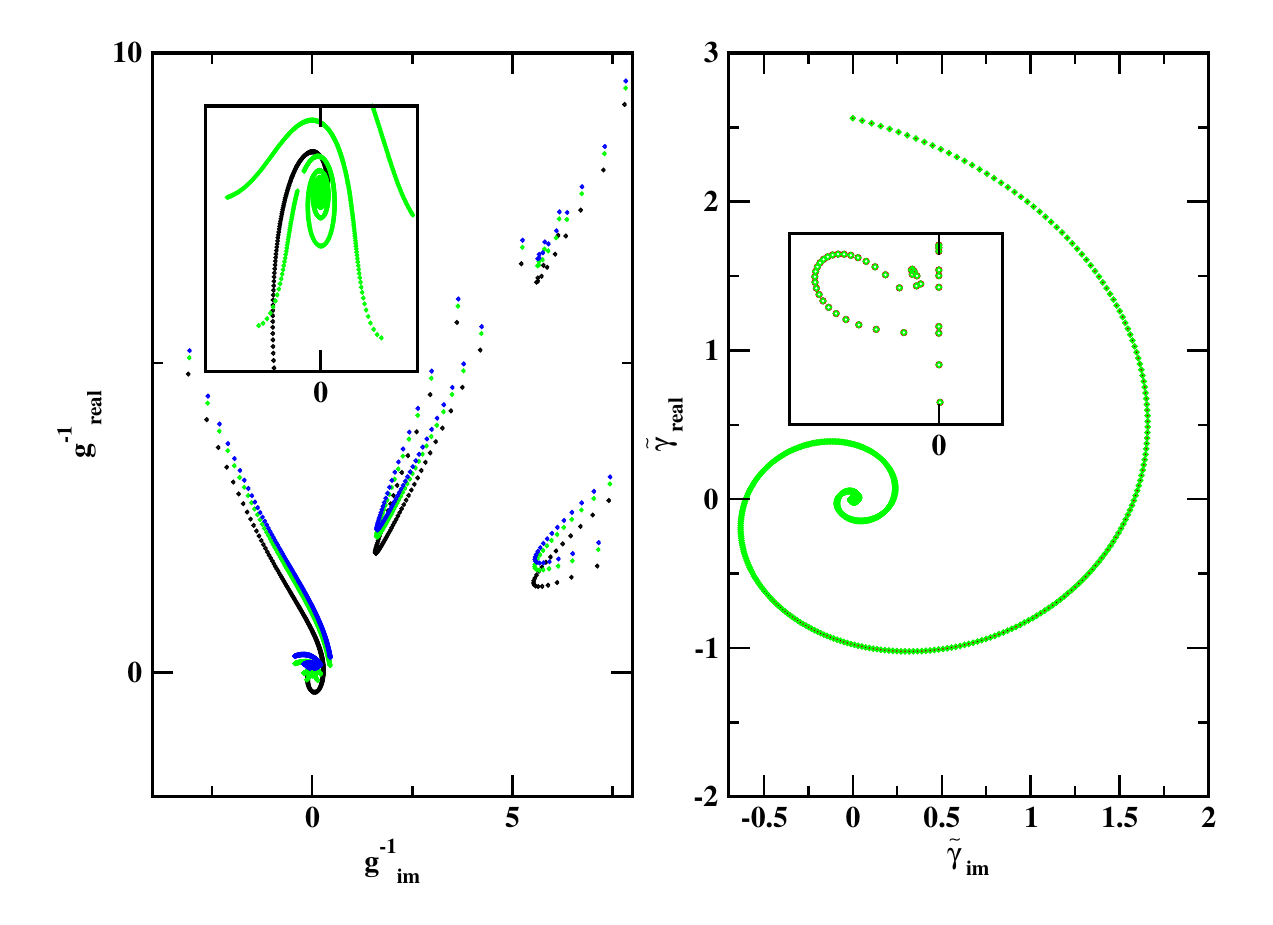}
     \caption{Left pannel is the renormallization of real part of correlation function ($\mathcal{G}^{-1}$ which is labeled as $g^{-1}$ in the plot) to imaginary part in the Matsubara space, Right pannel is the renormallization of real part of complex interaction with the imaginary part. We see existence of limit cycle in the interaction plane but the limit cycle in correlation function can be seen only when the bare interactions satisfy the condition $E_0\leq \gamma_0$.}
    \label{fig:flow_SO7}
\end{figure}
In order to investigate the whether there exist a critical regime for the model we considered computing the temperature scale with renormalized correlation function, which are of physical interest.
\begin{figure}
    \centering
    \includegraphics[scale=.8]{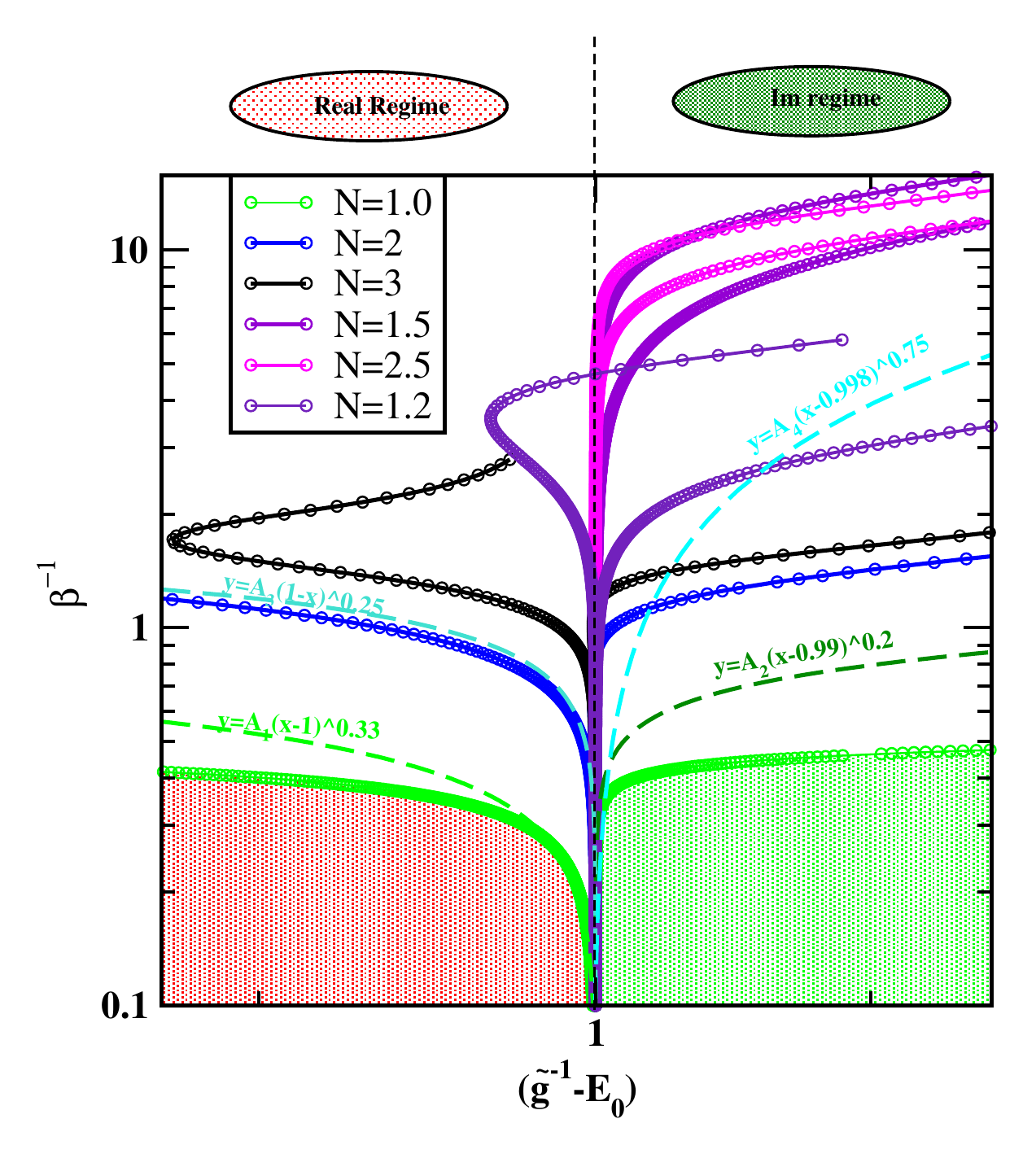}
     \caption{After doing the m-sum for $\mathcal{G}^{-1}$ we kept $\gamma=0.5$ and ground state energy $E_0=1.0$ constant and varying the power N of the complex potential, we got this phase diagram. The dashed line shows the fitting of the scales to power-law, which confirms the existence of a critical point for the integer value of N, particularly for N=1,2,3, and various fractional powers do not show any transition.}
      \label{fig:phased}
\end{figure}
It is evident from the left-right RG phase-diagram Figure \ref{fig:phased} is asymmetric so that the power law is not the same on either side as we have seen in conventional RG. This indicates that the modified RG renormalizes the imaginary phase of the BEC. The left part of the phase diagram still recovers the exponent as $(\tilde{\mathcal{G}}^{-1}-E_0)\propto N^2$ so therefore when complex power of the potential 1,2 capture the exponent 2/3, 1/2 respectively. The non-Hermitian BEC will have very different criticality than the BEC with interaction terms even in one dimension.
\section{Conclusions}
It is interesting to note that mathematically one can show that dual-space RG captures the deviations from the unitary regime and renormalize the complex parameter of the generic complex model consisting of discrete symmetries over a lattice or in a continuous medium. If one can come up with a field decomposition respecting the symmetries, then it is possible to go beyond the positive definiteness of the action measure or the square integrability of the wave function. It also ensures that the RG suggested in work accurately captures the wave function phase. \\
Bose-Einstein condensation has been studied in the complex model recently, and there are a few recent observations\cite{obs1,obs,li2021bose} as well.
Complex interaction in the harmonic oscillator gives the Non-Hermitian BEC transition and various critical exponents for potential power. Largely we looked into the various methods to get exact solutions for the complex potentials and their conserved regime where the unitary dynamics persist. Coherent states can still be constructed with slight modification, and the left-right g-variables can be justified through that.The existence limit cycle in RG flows are noticed in the various work\cite{sarkar2011center,de1995quantum,mueller2004renormalization,glazek,errglaz,Hammer,braaten2004renormalization,martini2012limit,bulycheva2014limit} earlier.As the conventional case, the limit cycle signifies the existence of the discrete symmetry such as parity, time, etc., also more importantly about the bound states, for example, in the Efimov state\cite{hammer2011efimov}.\\
As mentioned above, the RG will renormalize the imaginary phase, and it introduces the asymmetry in the phase diagram. We can verify by taking various limits to recover the existing results, exponents. One can conclude when $N\leq 2$ expect BEC-like transition and other cases only real-imaginary transition with no power-law behavior, for example, the fractional powers. For any other asymptotic $N>2$ and fixed $\beta$, we only see the chaotic behavior in the renormalized correlation function and complex potential. This we can also confirm from limit cycles of RG in the Left-Right case.



\ack
We wish to acknowledge for excellent research environment, research facilities of JNCASR.

\section*{Data Availability Statement}
All the data of the article is available on the main paper; any further information about the calculations can be discussed through the correspondence email provided in the article.

\appendix
\section{Detailed calculations on each sections}
\subsection{Frobenius Solution for coefficients}
We did not discuss the coefficients of the complex ODE in the beginning; although the complex extensions of the Frobenius theory exist for analytic functions the convergence does not guarantee always.
\begin{equation}
    \begin{split}
        c_{n+2}(n+1)(n+2)e^{i2\theta}+2(n-2N)c_{n-2N}e^{-i2N\theta}
        -c_{n-2}e^{-2iN\theta} \\+\frac{c_{n-4N-2}}{(2N+1)^2}e^{-i(4N+2)\theta}-(i\gamma)^{2N}c_{n-2N}-Ec_n=0
    \end{split}
\end{equation}
This yields the recurrence relation as ,
\begin{equation}
    \begin{split}
        c_{n+2}+c_{n+2N}\frac{n-2N}{(n+1)(n+2)}e^{-2(N+1)\theta} \\
        =\frac{c_{n-2}e^{-2iN\theta}-\frac{c_{n-4N-2}}{(2N+1)^2}e^{-i(4N+2)\theta}+(i\gamma)^{2N}c_{n-2N}+Ec_n}{(n+1)(n+2)}\\
      for \hspace{2mm} N\to 0,\gamma \to 0 \hspace{2mm} c_{n+2}=c_n\frac{E-n}{(n+1)(n+2)}
    \end{split}
\end{equation}
This turns out we do have the convergence for coefficients in special values of N but asymptotically if we do a convergence test it is hard to achieve from the method,
\begin{equation}
    \begin{split}
        \lim_{N\to \infty}\bigg|\frac{c_{n+2N}}{c_{n-2N}}\bigg|=\lim_{N\to\infty}\bigg|\frac{n-2N}{(i\gamma)^{2N}}e^{-2i\theta}\bigg|
    \end{split}
\end{equation}
\subsection{Conventional Calculations for Oscillator}
Action for oscillator (irrespective of fermions or bosons {\color{red}ref}) can be written as bosonic Grassmann integral as follows,
\begin{equation}
    \begin{split}
        \mathcal{S}=\int dt \bar{\phi}(i\frac{\partial}{\partial t}+\omega)\phi+E_{k} 
    \end{split}
\end{equation}

Scaling the above action by the regulator we get the following,
\begin{equation}
    \begin{split}
        \Gamma_{k}&=\mathcal{S}+\mathcal{S}_{k}\\
        &=\int dt \bar{\phi}(i\frac{\partial}{\partial t}+\omega)\phi+\bar{\phi}_{k}(R_{k})\phi_{k}+E_{k}
    \end{split}
\end{equation}
The path integral for the above action can be written as the $\mathcal{Z}_{k}=\int d[\bar{\phi}\phi]e^{\Gamma_{k}}$ in the usual field operators and the  
It can be derived the flow equation from above action which is known as Wetterich equation,
\begin{equation}
    \begin{split}
       e^{-\delta \Gamma_k}= \det(e^{-(\Gamma_k+R_k)^{-1}\delta R_k+(i\partial_t+R_k)^{-1}\delta R_k})\\
       =e^{-Tr(\Gamma_k+R_k)^{-1}\delta R_k+Tr(i\partial_t+R_k)^{-1}\delta R_k}
    \end{split}
\end{equation}
where $\tilde{\Gamma}=\frac{\delta^2 \Gamma}{\delta\phi_{k}\delta\phi_{k}}$
\begin{equation}
    \begin{split}
        \partial_{log k}\Gamma_{k}&=\frac{1}{2}Tr[(\tilde{\Gamma}_{k}+R_{k})^{-1}\partial_{log k}R_{k}] -\frac{1}{2}Tr[(i\frac{\partial}{\partial t}+R_{k})^{-1}\partial_{log k}R_{k}]
    \end{split}
\end{equation}
We choose a simple choice of the regulator  as $R_{k}(z)=(k^2-z)\theta(k^2-z)$ for conventional calculations later we show the dependence of the regulator on the effective action in non-Hermitian case.We get the following flow equation,
\begin{equation}
    \begin{split}
        &\int dt \partial_{log k}V_{k}=\frac{1}{2}\int dt \int \frac{dE}{2\pi}\theta(k^2-E^2)2k^2\big[\frac{1}{k^2+\omega^2}-\frac{1}{k^2}\big]\\
        &\partial_{k}E_{k}=\frac{1}{\pi}\frac{-\omega^2}{k^2+\omega^2}\implies E_{k}=-\frac{\omega}{\pi}arctan(\Lambda/\omega)\\
        &\Lambda \to \infty \implies E_{k=0}=\frac{\omega}{2}
     \end{split}
\end{equation}
\textbf{\textit{Sanity check}}: Consider the perturbation correction for $V'=\delta x$ in the above and solving the flow equation we get the following,
\begin{equation}
    \begin{split}
        E_{k}=\frac{\omega}{\pi}\sqrt{1-\big(\frac{\delta^2}{\omega^2}\big)}arctan\bigg(\frac{\Lambda}{\omega\sqrt{1-\big(\frac{\delta^2}{\omega^2}\big)}}\bigg)\hspace{2mm}
        As \hspace{2mm} \Lambda \to \infty ;\\
        E_{k}\to \frac{\omega}{2}\sqrt{1-\big(\frac{\delta^2}{\omega^2}\big)}=\frac{\omega}{2}\bigg(1-\frac{1}{2}\frac{\delta^2}{\omega^2} ...\bigg)=\frac{\omega}{2}-\frac{\delta^2}{4\omega}
    \end{split}
\end{equation}
Hence we can see from the above analysis as cutoff tending to large value we recover the second order perturbation theory result as $\langle(E_k)^{(2)}\rangle=-\frac{\delta^2}{4\omega}$ from the FRG. This also tells us the RG flows will be \textbf{\textit{parabolic}}  or in general conic sections in $E_k$ and $\delta$ plane.

\subsection{Conventional calculation for non-Hermitian oscillator}
\begin{figure}
    \centering
    \includegraphics[scale=1.0]{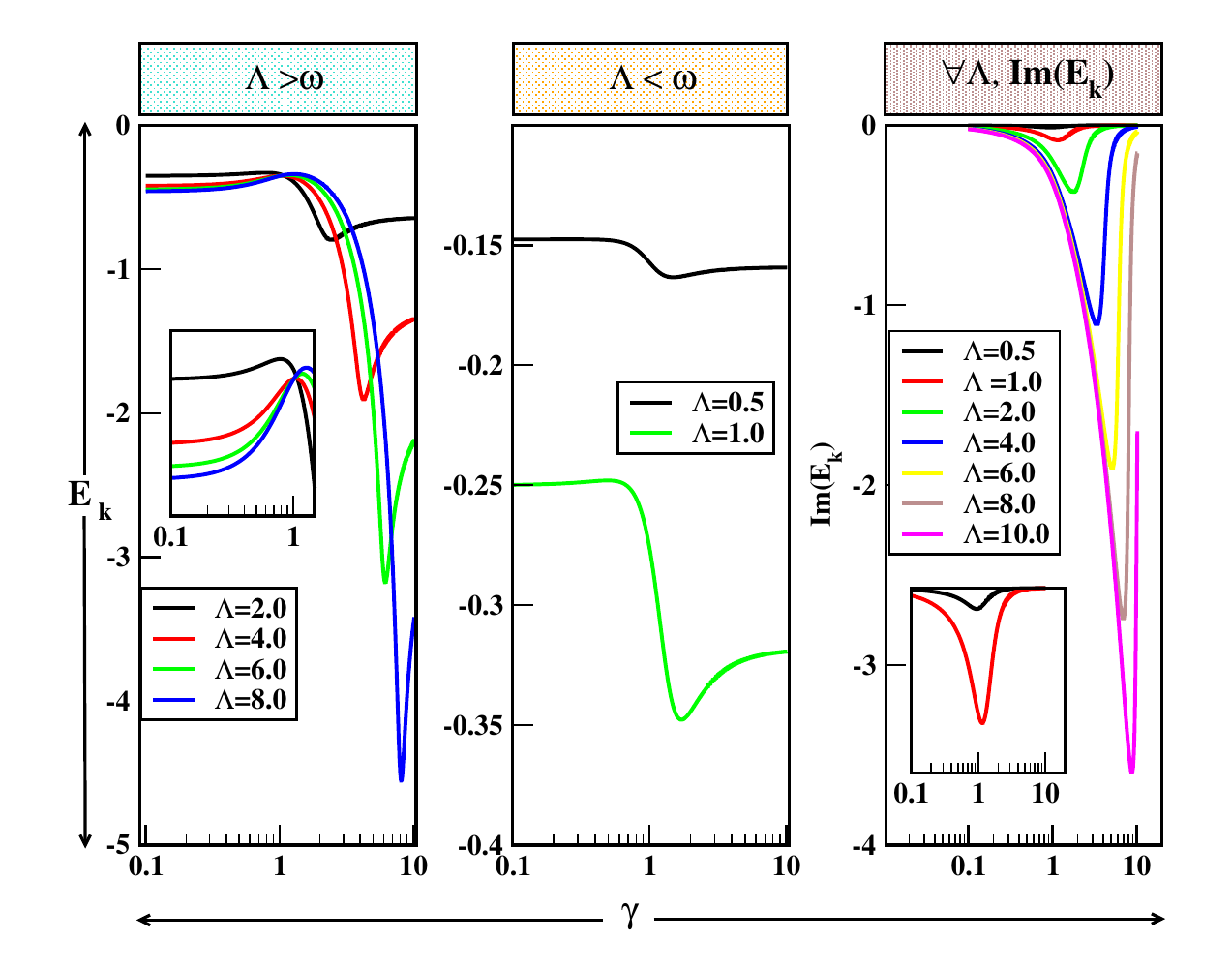}
     \caption{The panel labelled $\Lambda>\omega$ is the renormallized ground state energy for conventional regulator for various cuttoff scales for all the data oscillator frequency $\omega=1.0$ hence the  zero point energy $E_0=\frac{1}{2}$   }
    \label{fig:flow_ek}
\end{figure}

The FRG calculations which we performed in the earlier section will be extended to complex potential, and Here we derive flow equations for Non-Hermitian oscillator with various complex potentials.

\begin{equation}
    \begin{split}
         \mathcal{S}&=\int dt \bar{\phi}(i\frac{\partial}{\partial t}+\omega)\phi+E_{k} +\mathcal{S}_{im}\\
         &=\int dt \bar{\phi}(i\frac{\partial}{\partial t}+\omega)\phi+E_{k} +i\gamma(\bar{\phi}+\phi)
    \end{split}
\end{equation}
We can integrate out the variables and get following RG equations.
\begin{equation}
    \begin{split}
        \partial_{\log{k}} E_{k}=Re(I),\hspace{2mm}\partial_{\log{k}} \gamma_{k}=Im(I)\\
        E_{k}=\sqrt{\omega^2-\gamma^2+i\gamma\omega}\bigg(\arctan{\frac{\Lambda}{\sqrt{\omega^2-\gamma^2+i\gamma\omega}}}\bigg)
    \end{split}
\end{equation}
$I=\bigg(\int^{\Lambda}_{0} \frac{\omega^2-\gamma^2+i\gamma\omega}{k^2+\omega^2-\gamma^2+i\gamma\omega}dk\bigg)$
Renormalized $E_k$ with various parameter regime are plotted in the \ref{fig:flow_ek}. We notice that the higher cutoff will increase the imaginary weight of the $E_k$ as complex interaction increases and the asymmetry in the weight also increases as $\gamma>\omega=1$.This clearly indicate that we are in the imaginary eigenvalue regime. 
The renormallization of bosonic correlation function for the diagonal regulator FRG and 1-loop RG  the non linearity is second order can be solved analytically and compared. We can also perform the Matsubara sum to get characteristic temperature scale.Here we considered the same order coupling polynomial in RG-ODE to compare the solutions with the other methods.
\begin{equation}
    \begin{split}
        -\frac{\partial}{\partial\tau_n}\tilde{\mathcal{G}}_{n}= \mathcal{G}^2 -\gamma^2, \hspace{2mm}
         -\frac{\partial}{\partial\tau_n}\tilde{\gamma}_{n}= \gamma\mathcal{G},\hspace{2mm}\implies
         \frac{\partial\tilde{\mathcal{G}}_{n}}{\partial\gamma}= \frac{\mathcal{G}^2 -\gamma^2}{\gamma\mathcal{G}}
    \end{split}
\end{equation}
\begin{equation}
    \begin{split}
        \frac{(\mathcal{G}^{-1})^2}{\gamma^2}=\log{\frac{1}{\gamma}}+C
    \end{split}
\end{equation}
We can see the how the Matsubara time which is flow parameter in our case  scales with the dissipation or non Hermitian strength$(\gamma)$
\begin{equation}
    \begin{split}
        \tilde{\mathcal{G}}^{-1}=\oint \frac{1}{z-E}\frac{\beta}{1-e^{-\beta z}}dz=\gamma\sqrt{\log{\frac{C}{\gamma}}}\\
       \langle n_B\rangle =\frac{1}{1-e^{-\beta E}}=\gamma\sqrt{\log{\frac{C}{\gamma}}}\\
       \therefore T_{B}=\frac{E}{2\pi\log{\big(1-\frac{\gamma}{\log{\frac{C}{\gamma}}}\big)}}
    \end{split}
\end{equation}
Matsubara time flow with dissipation gets the following form at one loop correction.
\begin{equation}
    \begin{split}
        \tau=\int \frac{d\gamma}{\gamma^2\sqrt{\log{\frac{C}{\gamma}}}}
        =\sqrt{\pi}Erfi(\log{\frac{C}{\gamma}})+A
    \end{split}
\end{equation}
\subsection{Functional Renormalization calculations}
The derivation for the flow is derived in the main article here we do some contour integrals to check some special limits.We will consider the potential from now on as $(ix)^{2N}$, which is handy in regularization and integrals to calculate ground state.
Momentum scales the potential as the $U \propto k^2$ in conventional case, Hence complex potential can be scaled as $(\gamma/\omega)^{1/2N}e^{i N\pi/2}k^{2N}$ hence we can modify the flow equations as follows,

\begin{equation}
    \begin{split}
        &\dot{\Gamma}=\frac{1}{2}\bigg(\frac{k^2}{k^2+\omega^2+(\gamma/\omega)^{1/2N}e^{i N\pi}k^{2N}}\bigg)-\frac{1}{2}\\
        &U_{eff}=-\int^{\Lambda}_0\bigg(\frac{\omega^2+(\gamma/\omega)^{1/2N}e^{i N\pi}k^{2N}}{k^2+\omega^2+(\gamma/\omega)^{1/2N}e^{i N\pi}k^{2N}}\bigg)dk\\
        &\omega^2+(\gamma/\omega)^{1/2N}e^{i N\pi}k^{2N}=\Omega \\ &\implies 2N(\gamma/\omega)^{1/2N}e^{i N\pi/2}k^{2N-1}dk=d\Omega\\
        &U_{eff}=i\int^{\omega^2+\big(\frac{\gamma}{\omega}\big)^{1/2N}e^{i\pi N}\Lambda^{2N}}_{\omega^2}\mathcal{M}(\frac{\gamma}{\omega}\big)^{1/4N^2}d\Omega\\
        &\mathcal{M}=\bigg(\frac{\Omega}{e^{-i\pi}(\frac{\gamma}{\omega})^{1/2N}(\Omega-\omega^2)^{1/N}+\Omega}\bigg)(\Omega-\omega^2)^{(1-2N)/2N}\\
         &\lim_{N\to \infty,\omega\to 1} U_{eff}=i\big( (1+e^{iN\pi}\Lambda^{2N})^2/2-1/2\big)
    \end{split}
\end{equation}
If we now use the regulator in the two halves of the momentum vector for positive and negative, we choose opposite phase for each of the vectors,
\begin{equation}
    \begin{split}
       U_{eff}=-\int^{\Lambda}_0\bigg(\frac{\omega^2+\big(\frac{\gamma}{\omega}\big)^{\frac{1}{2N}}e^{i\pi N}k^{2N}}{k^2+\omega^2+\big(\frac{\gamma}{\omega}\big)^{\frac{1}{N}}e^{i\pi N}k^{2N}}\bigg)dk +\int^{\Lambda}_0\bigg(\frac{\omega^2+\big(\frac{\gamma}{\omega}\big)^{\frac{1}{2N}}e^{-i\pi N}k^{2N}}{k^2+\omega^2+\big(\frac{\gamma}{\omega}\big)^{\frac{1}{2N}}e^{-i\pi N}k^{2N}}\bigg)dk\\
       =-\frac{N}{2}\oint \frac{Ze^{-\frac{i\pi}{2}}\big(\gamma/\omega\big)^{\frac{1}{4N^2}}(Z-\omega^2)^{(1-2N )/2N}}{(Z-\omega^2)^{\frac{1}{N}}e^{-i\pi}(\frac{\gamma}{\omega})^{1/2N^2}+Z} dZ   +\frac{N}{2}\oint \frac{Z^*e^{\frac{i\pi}{2}}\big(\gamma/\omega\big)^{\frac{1}{4N^2}}(Z^*-\omega^2)^{(1-2N )/2N}}{(Z^*-\omega^2)^{\frac{1}{N}}e^{i\pi}(\frac{\gamma}{\omega})^{1/2N^2}+Z^*} dZ^*\\
       where \hspace{2mm} Z=\omega^2+\big(\frac{\gamma}{\omega}\big)^{\frac{1}{2N}}e^{i\pi N}k^{2N} 
       k=(Z-\omega^2)^{\frac{1}{2N}}e^{-i\pi/2}(\frac{\gamma}{\omega})^{1/4N^2}\\ \implies dk=\frac{N}{2}e^{-\frac{i\pi}{2}}\big(\gamma/\omega\big)^{\frac{1}{4N^2}}(Z-\omega^2)^{(1-2N )/2N}dZ
    \end{split}
\end{equation}
The regulator for the positive and negative momentum with general phase can be analyzed separately as the following,
\begin{equation}
    \begin{split}
        \lim_{\omega\to 0} U_{eff}= \int^{\Lambda}_{0} \bigg(\frac{e^{\frac{1}{2} i \pi  (2 N-1)} k^{2 N}}{1+e^{\frac{1}{2} i \pi  (2 N-1)} k^{2 N}}-\frac{e^{\frac{1}{2} (-i) \pi  (2 N-1)} k^{2 N}}{1+e^{\frac{1}{2} (-i) \pi  (2 N-1)} k^{2 N}}\bigg) \, dk\\
        =\Lambda \, _2F_1\left(1,\frac{1}{2 N};1+\frac{1}{2 N};-e^{-\frac{1}{2} i (2 N-1) \pi } \Lambda^{2 N}\right)\\-
        \Lambda\, _2F_1\left(1,\frac{1}{2 N};1+\frac{1}{2 N};-e^{\frac{1}{2} i (2 N-1) \pi } \Lambda^{2 N}\right)\\
        _2F_1 \to Gauss-Hypergeometric Function
    \end{split}
\end{equation}
This basically compares well with the Left-right RG for the action.
For $\omega \neq 0$ and for $N=1$ we have the following ground state energy.
\begin{equation}
\label{eq:genericE}
    \begin{split}
      U_{eff}= \frac{1}{2} \gamma e^{\frac{i \pi }{2}} \log \left(\gamma e^{\frac{i \pi }{2}} \Lambda+\Lambda^2+\omega ^2\right)-
      \frac{\left(-2 \omega ^2+\gamma^2 e^{i \pi }\right) \tan ^{-1}\left(\frac{2 \Lambda+\gamma e^{\frac{i \pi }{2}}}{\sqrt{4 \omega ^2-\gamma^2 e^{i \pi }}}\right)}{\sqrt{4 \omega ^2-b^2 e^{i \pi }}}\\
      for \hspace{2mm} 2N=2; \hspace{2mm}\to U_{eff}=\frac{\omega  \tan ^{-1}\left(\frac{\Lambda \sqrt{1+\gamma e^{i \pi }}}{\omega }\right)}{\left(1+\gamma e^{i \pi }\right)^{3/2}}+\frac{\gamma e^{i \pi } \Lambda}{1+\gamma e^{i \pi }} 
    \end{split}
\end{equation}
As  $\gamma\to0$  above expressions \ref{eq:genericE} reduce to the conventional FRG expression.
For exponential regulator $R_k=\frac{k^2}{e^{\frac{k^2}{\Lambda^2}}-1}$ we find it hard to solve analytically due to the complex interaction but in some limits we still can expect the above result for low $\omega$ also it is strongly dependent on cutoff scale, which we don't discuss here as it involves slightly different approach to problem.Our aim to show the functional integration in tau space does the very similar job that of the modified regulator but it has more control of the order of perturbation in the normal ordered operator product.

\subsection{Calculation of RG equation in $\tau$-space}
We show the explicit calculation for the imaginary time RG, Let's start from the the action for the conventional basis.
\begin{equation}
    \begin{split}
        S&=\sum^{\beta/2}_{n=-\beta/2}\bar{\phi}_n(\mathcal{G}^{-1})\phi_n+(i\gamma)^N(\bar{\phi}_n+\phi_n)^N\\
        &S=\sum^{n=-1}_{n=-\frac{\beta}{2}}S_n +S_0+\sum^{n=\frac{\beta}{2}}_{n=1}S_n\\
        &S=\sum^{n=0}_{n=-\frac{\beta}{2}-1}S_{n+1}+S_0+\sum^{n=\frac{\beta}{2}+1}_{n=0}S_{n-1}
    \end{split}
\end{equation}
after integrating out the pair of variables we can show that,
\begin{equation}
    \begin{split}
        S_{eff}=S-S_{0}+\sum_{n}S_{n+1}S^{-1}_{0}S_{n-1}\\
        \therefore \delta S =S_{eff}-S=-S_0+\sum_{n}S_{n+1}S^{-1}_{0}S_{n-1}\\
        \delta S =-S_0 -S^{2}_{n-1}S^{-1}_{0}+SS^{-1}_{0}S_{n-1}
    \end{split}
\end{equation}
leading terms in one loop general-RG equation towards $\tau \to \frac{-\beta}{2}$,
\begin{equation}
    \begin{split}
    \tau \to \frac{-\beta}{2}  \hspace{2mm}  S_0 \delta S=-(S^2_0-S^2_{n-1}+cross-terms)\\
     \tau \to \frac{\beta}{2}  \hspace{2mm}    S_0 \delta S=-(S^2_0+S^2_{n+1}+cross-terms)
    \end{split}
\end{equation}
Where $S_0$ is the following,
\begin{equation}
    \begin{split}
      S_0=\bar{\phi}_0(\mathcal{G}^{-1})\phi_0+(i\gamma)^N(\bar{\phi}_0+\phi_0)^N
    \end{split}
\end{equation}
We use the binomial expansion in normal ordered operators as the following,
\begin{equation}
    \begin{split}
      S_0=\bar{\phi}_0(\mathcal{G}^{-1})\phi_0+(i\gamma)^N\sum^{N}_{m=0}\sum^{min(m,N-m)}_{l=0}\bar{\phi}^{m-l}_0\phi^{N-m-l}_0
    \end{split}
\end{equation}
In bosonic grassmann we can show the following
\begin{equation}
    \begin{split}
      \mathcal{S}_{int} = (i\gamma)^{2N}\sum^{N}_{m=0}\sum^{min(m,N-m)}_{l=0}\bar{\phi}^{m-l}_0\phi^{N-m-l}_0 \sum^{N}_{m'=0}\sum^{min(m',N-m')}_{l'=0}\bar{\phi}^{N-m'-l'}_0\phi^{m'-l'}_0\\
      = (i\gamma)^{2N}\sum^{N}_{m,m'=0}\sum^{min(m,\frac{N}{2}-m)}_{l=0}\sum^{min(m',N-m')}_{l'=0} \bar{\phi}^{N+m'-m-l-l'}_0\phi^{N-m+m'-l-l'}_0\delta_{m,m'}\delta_{l,l'}\\
      =  (i\gamma)^{2N}N\bar{\phi}^{N-2l}_0\phi^{N-2l}_0 \end{split}
\end{equation}
Above can contribute to the leading correction to RG equations when $m-l=N-m-l$ which implies that $N=2m$\\

Contributions to RG recurrence equations by expanding the action with the commutation algebra these following summutions we get and by shifting the coefficients we can find the operator structures as well,
\subsubsection{contribution to $\tilde{\mathcal{G}}^{-1}_0$ Renormalization}
If we compute the vertex correspond to imaginary interaction which appear in the action expansion, we get the following,
\begin{equation}
    \begin{split}
        \frac{1}{2}(i\gamma)^N\Bigg( \sum^{N}_{m=0}\sum^{min(m,N-m)}_{l=0}\begin{Bmatrix}N\\m\end{Bmatrix}_{l}\bar{\phi}^{m-l}\phi^{N-m-l}+
        \sum^{N}_{m=0}\sum^{min(m,N-m)}_{l=0}\begin{Bmatrix}N\\m\end{Bmatrix}_{l}\bar{\phi}^{N-m-l}\phi^{m-l}\Bigg)\\
        S_{eff} \approx (i\gamma)^{2N}\sum^{N}_{m=0}\sum^{min(m,N-m)}_{l=0}\begin{Bmatrix}N\\m\end{Bmatrix}_{l} \bar{\phi}^{N-2l-1}\phi^{N-2l-1} \hspace{2mm} \forall N \ge 1
    \end{split}
\end{equation}
This show the leading contribution will be for $\mathcal{G}^{-1}_0$ is $(i\gamma)^{2N}$
\subsubsection{contribution to $\gamma$ Renormalization}
This correspond to the vertex that appear between the cross terms of real and imaginary interaction,
\begin{equation}
    \begin{split}
        (i\gamma)^N \mathcal{G}^{-1}_0\sum^{N}_{m=0}\sum^{min(m,N-m)}_{l=0}\begin{Bmatrix}N\\m\end{Bmatrix}_{l}\bar{\phi}^{m-l+1}\phi^{N-m-l+1} \\
        (i\gamma)^N \mathcal{G}^{-1}_0\sum^{N}_{m=0}\sum^{min(N-m,m)}_{l=0}\begin{Bmatrix}N\\m\end{Bmatrix}_{l}\bar{\phi}^{N-m-l+1}\phi^{m-l+1}
    \end{split}
\end{equation}
After integrating out the pair of variables  we can show easily this always contributes to the $\gamma$. Now these similar calculations can be done on complex time and show the RG equations are as follows,
\begin{equation}
    \begin{split}
        \mathcal{G}^{-1}_n=\mathcal{G}^{-1}_{n-1}-(\mathcal{G}^{-1}_n)^2 +(i\gamma_n)^{2N}\\
        (i\gamma_n)^N=(i\gamma_{n-1})^N+(i\gamma)^N\mathcal{G}^{-1}_n  
    \end{split}
\end{equation}
\subsection{Two-body Term from the Schrodinger Equation \ref{eq:schroo}}
We can see the various two body terms after the analytic continuation of g-numbers, If we find a condition when these imaginary interactions vanish we can formally construct the path-integral in the a-continued coherent states.
\begin{equation}
    \begin{split}
        \sum_{\alpha\beta\gamma\delta}\langle\alpha\beta|\tilde{V}|\gamma\delta\rangle\tilde{\phi}^{*}_{\alpha}\tilde{\phi}^{*}_{\beta}\frac{\partial}{\partial \tilde{\phi}^{*}_{\gamma}}\frac{\partial}{\partial \tilde{\phi}^{*}_{\delta}}+
         \sum_{\alpha\beta\gamma\delta}\langle\alpha\beta|\tilde{V}|\gamma\delta\rangle(\phi^{*}_{\alpha}-i\eta^{*}_{\alpha})(\phi^{*}_{\beta}-i\eta^{*}_{\beta})\bigg(\frac{\partial}{\partial \phi^{*}_{\gamma}}\frac{\partial}{\partial \phi^{*}_{\delta}}-\\
         \frac{\partial}{\partial \eta^{*}_{\gamma}}\frac{\partial}{\partial \eta^{*}_{\delta}}-i \frac{\partial}{\partial \phi^{*}_{\gamma}}\frac{\partial}{\partial \eta^{*}_{\delta}}-i\frac{\partial}{\partial \eta^{*}_{\gamma}}\frac{\partial}{\partial \phi^{*}_{\delta}}\bigg)\\
         \sum_{\alpha\beta\gamma\delta}\langle\alpha\beta|\tilde{V}|\gamma\delta\rangle(\phi^{*}_{\alpha}\phi^{*}_{\beta}-i\eta^{*}_{\alpha}\phi^{*}_{\beta}-i\eta^{*}_{\beta}\phi^{*}_{\alpha}-\eta^{*}_{\alpha}\eta^{*}_{\beta})\bigg(\frac{\partial}{\partial \phi^{*}_{\gamma}}\frac{\partial}{\partial \phi^{*}_{\delta}}\\
         -\frac{\partial}{\partial \eta^{*}_{\gamma}}\frac{\partial}{\partial \eta^{*}_{\delta}}-i \frac{\partial}{\partial \phi^{*}_{\gamma}}\frac{\partial}{\partial \eta^{*}_{\delta}}-i\frac{\partial}{\partial \eta^{*}_{\gamma}}\frac{\partial}{\partial \phi^{*}_{\delta}}\bigg)
    \end{split}
\end{equation}
we can collect the complex terms explicitly arising because of the analytic continuation from the above full two body term as the following,
\begin{equation}
\label{eq:two_body}
    \begin{split}
       im(T_{\alpha\beta\gamma\delta})= -i\eta^{*}_{\alpha}\phi^{*}_{\beta}\frac{\partial}{\partial \phi^{*}_{\gamma}}\frac{\partial}{\partial \phi^{*}_{\delta}} +i\eta^{*}_{\alpha}\phi^{*}_{\beta}\frac{\partial}{\partial \eta^{*}_{\gamma}}\frac{\partial}{\partial \eta^{*}_{\delta}}-i\eta^{*}_{\beta}\phi^{*}_{\alpha}\frac{\partial}{\partial \phi^{*}_{\gamma}}\frac{\partial}{\partial \phi^{*}_{\delta}}+i\eta^{*}_{\beta}\phi^{*}_{\alpha}\frac{\partial}{\partial \eta^{*}_{\gamma}}\frac{\partial}{\partial \eta^{*}_{\delta}}\\
        -i\phi^{*}_{\alpha}\phi^{*}_{\beta} \frac{\partial}{\partial \phi^{*}_{\gamma}}\frac{\partial}{\partial \eta^{*}_{\delta}}-i\phi^{*}_{\alpha}\phi^{*}_{\beta}\frac{\partial}{\partial \eta^{*}_{\gamma}}\frac{\partial}{\partial \phi^{*}_{\delta}}+i\eta^{*}_{\alpha}\eta^{*}_{\beta} \frac{\partial}{\partial \phi^{*}_{\gamma}}\frac{\partial}{\partial \eta^{*}_{\delta}}+i\eta^{*}_{\alpha}\eta^{*}_{\beta}\frac{\partial}{\partial \eta^{*}_{\gamma}}\frac{\partial}{\partial \phi^{*}_{\delta}}\\
        im(T_{\alpha\beta\gamma\delta})\psi=0 \hspace{2mm} iff \hspace{2mm} \psi=e^{k\int \phi^*_\alpha  d\phi^*_\beta}\xi_0 \otimes \chi_0e^{-k\int \eta^*_\alpha d\eta^*_\beta}
    \end{split}
\end{equation}
Various permutation of the coherent states will introduce the certain interaction for example if we take the wave function which derived for the single body term as $\psi=e^{k\int \phi^*_\alpha  d\phi^*_\beta}\xi_0 \otimes \chi_0e^{-k\int \eta^*_\alpha d\eta^*_\beta}$ will take all the terms in \ref{eq:two_body} to zero. This exercise shows that to preserve symmetry, one needs the product states. Also, if it happens, so we get common many-body eigenstate for one body and two-body terms of Hamiltonian.The quantity in the product state $i\big(\int \phi^*_{\alpha}d\phi^*_{\beta}-\int \eta^*_{\alpha}d\eta^*_{\beta}\big)\neq n\pi$ will lead to the anyonic coherent states which is discussed somewhere else\cite{subramanyan2019correlations,grundberg1995coherent}.These arbitrary phase-coherent states can be worked out in the path integral with shifted ground state energy.
\section*{References}
\bibliographystyle{unsrt}
\bibliography{ref.bib}

\end{document}